\documentclass[aps,preprint,amsmath,amssymb,showkeys,superscriptaddress]{revtex4-2}
\usepackage[utf8]{inputenc}

\usepackage{graphicx}
\usepackage{dsfont}
\usepackage{booktabs}
\usepackage{enumitem}
\usepackage{tcolorbox}
\usepackage{tikz}
\usepackage{hyperref}
\usepackage{tcolorbox}

\renewcommand{\vec}[1]{\boldsymbol{#1}} 

\newcommand{\gv}[1]{\ensuremath{\mbox{\boldmath$ #1 $}}} 

\newcommand{\bdot}{\!\cdot\!}

\newcommand{\diff}{\mathrm{d}}
\renewcommand{\d}[2]{\frac{\diff #1}{\diff #2}} 
\newcommand{\dd}[2]{\frac{\diff^2 #1}{\diff #2^2}} 
\newcommand{\pd}[2]{\frac{\partial #1}{\partial #2}}

\newcommand{\pdd}[2]{\frac{\partial^2 #1}{\partial #2^2}}

\newcommand{\pdx}[1]{\frac{\partial}{\partial #1}} 

\newcommand{\grad}[1]{\gv{\nabla} #1} 

\renewcommand{\div}[1]{\gv{\nabla} \bdot #1} 
\newcommand{\mean}[1]{\left< #1 \right>}
\newcommand{\laplacian}[1]{\nabla^2 #1}

\newcommand{\pdt}[1]{\partial_t #1}

\newcommand{\ddx}{\d{}{x}}
\newcommand{\dddx}{\dd{}{x}}

\newcommand{\Biot}{Bi}
\newcommand{\Peclet}{\mathrm{Pe}}
\newcommand{\Pecletlocal}{\mathrm{Pe}_{\overline{u}}}
\newcommand{\Reynolds}{\mathrm{Re}}
\newcommand{\Rayleigh}{\mbox{\textit{Ra}}}

\newcommand{\evec}{\hat{\vec e}}

\begin{document}
\title{Thermo-viscous instability of flow in a weakly heat-conducting channel}

\author{Federico Lanza}
\email{federico.lanza.5@gmail.com} 
\affiliation{PoreLab, The Njord Centre, University of Oslo, Norway}

\author{Gaute Linga}
\affiliation{PoreLab, The Njord Centre, University of Oslo, Norway}
\affiliation{PoreLab, Department of Physics, Norwegian University of Science and Technology, Norway}

\author{Fabian Barras}
\affiliation{PoreLab, The Njord Centre, University of Oslo, Norway}

\author{Eirik G.\ Flekkøy}
\affiliation{PoreLab, The Njord Centre, University of Oslo, Norway}
\affiliation{PoreLab, Department of Chemistry, Norwegian University of Science and Technology, Norway}

\begin{abstract}
An instability may arise when a hot viscous fluid enters a thin gap and cools through heat transfer to a colder surrounding environment. Fluids whose viscosity increases strongly upon cooling create a positive feedback in which warmer regions flow faster and cool more slowly, leading to the formation of thermo-viscous "fingers". Here we investigate this mechanism in the long time, small Biot number regime, where cooling through the plates is weak but acts over sufficiently long times that the temperature becomes nearly uniform across the gap heat. This asymptotic limit enables a depth-averaged description that incorporates both thermal diffusion and hydrodynamic (Taylor) dispersion, allowing us to analyze the dependence of the instability on the Péclet number, viscosity contrast, and wall cooling rate. Using numerical simulations of temperature-dependent viscous flow in a Hele-Shaw geometry, we show that fingering instabilities emerge in response to small inlet perturbations within a range of Péclet numbers and viscosity contrasts. From linear stability analysis we find the dispersion relation and quantify how the fastest growth rate $\gamma_{\max}$ and corresponding wavenumber $k_{\max}$ depend on the global parameters. We further derive analytical expressions for $\gamma_{\max}$ and $k_{\max}$ in the limit of high Péclet number and large viscosity contrast, revealing the scaling behavior that controls pattern selection. These results clarify the physical mechanisms driving thermo-viscous fingering in the small Biot number regime and have implications for systems in which temperature-dependent viscous fluids are confined within narrow gaps, such as lubrication flows in mechanical components and magma invasion in small scale fissures.

\end{abstract}

\maketitle


\section{Introduction}\label{sec:introduction}
The flow of thermo-viscous fluids in confined geometries, such as fissures, porous media, or interfaces between contacting solids, is relevant to a variety of natural and industrial processes, such as subsurface magma flow, geothermal energy extraction \citep{Mcdowell2016}, enhanced oil recovery \citep{Austin-Adigio2020} as well as friction and lubrication between mechanical compounds \citep{Khonsari2017}. In these systems, the fluid viscosity may depend strongly on temperature, typically decreasing as the temperature increases. This temperature-viscosity coupling introduces a feedback mechanism: a local increase in temperature reduces the viscosity, which in turn promotes faster flow and enhanced heat transport in that region. As a result, small perturbations in the temperature or velocity field can become amplified, giving rise to the formation of finger-like structures and complex flow patterns.

A prominent example of thermo-viscous flow in a confined geometry is magma transport through the Earth's crust. The viscosity of magma can vary by several orders of magnitude in response to temperature changes of just a few hundred degrees Celsius \citep{Blatt2006}. Magma often flows through the Earth's crust via thin conduits, like fissures or dikes, whose thickness is much smaller than its other dimensions, effectively creating a quasi-planar geometry \citep{Jones2017, Pansino2019}. These sheet-like intrusions of magmatic rock often display finger-like structures \citep{Galland2019}. These regions of enhanced flow are typically referred to as channels or tubes, depending on their morphology and scale.

The prevalent interpretation of such finger patterns observed in the field is that they result from a Saffman-Taylor type instability \citep{SaffmanTaylor1958, homsy1987viscous}, which occurs when a less viscous magma displaces a more viscous host rock \citep{Pollard1975}. The combined effects of viscosity contrast and geometrical confinement make this a natural setting for the development of viscous fingering, further modulated by temperature-dependent rheology.

A related class of instabilities arises in fluids with concentration-dependent viscosity \citep{Tan1986, Tan1988, DeWit1999, Jha2011, DeWit2020}, where solute transport modifies the flow properties. In this case, the advection-diffusion equation for the solute concentration is formally analogous to that for temperature in thermo-viscous models, although key differences arise in the source terms, as thermal models typically include heat exchanges with boundaries, while solute transport may involve adsorption or reaction effects. These studies show that the coupling between concentration and viscosity can give rise to fingering instability dynamics analogous to those in thermo-viscous systems, with differences emerging from the specific transport mechanisms and boundary conditions.

Several experiments have reproduced thermo-viscous fingering in the laboratory using fluids with strong temperature-dependent viscosity contrasts, including paraffin wax \citep{Whitehead1991}, corn syrup \citep{Wylie1999}, glycerin \citep{Holloway2005, Nagatsu2009}, and polyethylene-glycol wax \citep{Anderson2005}. These studies showed that, for appropriate combinations of flow rate and temperature contrast, small perturbations in the inlet flow can trigger a fingering instability. In particular, \citet{Holloway2005} demonstrated that the early-time growth is exponential, and they identified a critical viscosity ratio between the injected hot fluid and the cooler fluid in equilibrium with the plates, below which instability occurs.

A number of theoretical and numerical studies have investigated thermo-viscous instability in quasi-planar geometries \citep{Helfrich1995, Wylie1995, Morris1996, Balmforth2004, Holloway2006, Diniega2013, Taylor-West2025}. Using a gap-averaged formulation and assuming high Péclet number, \citet{Helfrich1995} showed that instability is possible only when the viscosity ratio falls below a critical value that depends on the inlet conditions. Subsequent studies explored how the most unstable wavelength depends on the thermal entry length \citep{Wylie1995, Morris1996}, as well as on the cell gap width \citep{Holloway2006}.
A common assumption in these works is that the temperature at the interface between the fluid and the plates is taken to be constant in time. As heating of the solid plate is ignored, their analysis should be considered valid only for times less than the typical time for heat to diffuse through the system. For meter-wide fissures in basaltic rocks, this time is of the order of days, longer than the typical time for an instability to develop and lead to magma channelization, thus confirming the validity of this hypothesis in this context.

Moreover, it is important to note that fully resolved three-dimensional simulations and analyses, like the ones from \citet{Wylie1995} and \citet{Holloway2006}, inherently capture the spreading of heat (or solute) associated with dispersion, because they explicitly resolve the cross-gap temperature and velocity field. In contrast, gap-averaged models have to model the evolution of the cross-gap averaged fields, possibly taking into account the feedback between the cross-gap temperature and the velocity in a consistent way. In their pioneering work on thermal instabilities, \citet{Helfrich1995} did not consider variations in mobility across the gap, despite modeling a strongly non-uniform (sinusoidal) temperature profile. More refined cross-gap averaging techniques were introduced later, for example modeling a thermal boundary layer in contact with the cold wall evolving with time \citep{Balmforth2004, Taylor-West2025}.

In the present work, we revisit thermo-viscous fingering in the regime where out-of-plane heat transfer influences the wall temperature. In this long-time limit, heat has had sufficient time to diffuse through both the fluid and the solid plates, so that wall heating can no longer be neglected. Because the thermal resistance of the plates becomes comparable to, or larger than, that of the fluid layer, this regime corresponds to a small Biot number. This allows a cross-gap averaging that consistently captures the coupling between temperature and velocity. The resulting reduced model includes a contribution from Taylor dispersion \citep{Taylor1953, Aris1956}, which becomes relevant at sufficiently high Péclet numbers. Such long time, small Biot number conditions arise naturally in systems where fluids sensitive to temperature are confined within thin gaps, such as glycerol and mineral oils in lubrication films between mechanical components, or magma flowing through few millimeters-wide fissures. These applications are discussed in more detail in \S\,\ref{sec:discussion_applications}.

This article is organized as follows. In \S\,\ref{sec:model}, we introduce the governing equations and the base state solution. Section \ref{sec:full} presents results from time-dependent numerical simulations of the full nonlinear problem. The linear stability analysis is developed in \S\,\ref{sec:lsa}, where we also compare its predictions with the nonlinear simulation results. Finally, in \S\,\ref{sec:discussion}, we summarize and discuss the main findings.


\section{Fully coupled gap-averaged model}
\label{sec:model}

\begin{figure}
  \centering
    \begin{tikzpicture}[scale=1.0]
    \def\tagh{2.3cm}
    
    \node[] (A) at (0,0) {\includegraphics[width=0.56\columnwidth]{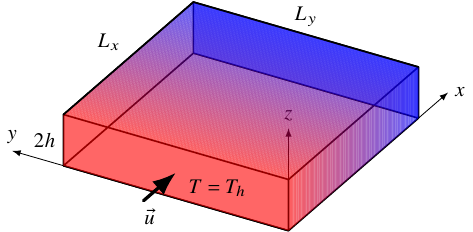}
    \includegraphics[width=0.38\columnwidth]{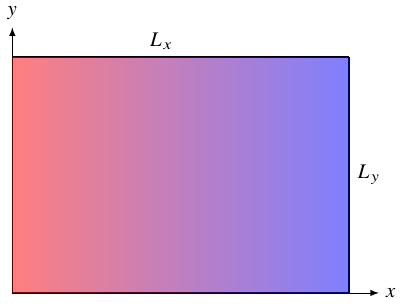}
    };
    \node[] (Atag) at (-6cm, \tagh) {{(\textit{a})}};
    \node[] (Atag) at (1.3cm, \tagh) {{(\textit{b})}};

    \end{tikzpicture}
  \caption{
  Schematic view of the model setup, with (\textit{a}) a full three-dimensional view and (\textit{b}) top view of the gap-averaged channel considered in the present work. Hot, less viscous fluid, shown in red, is injected into a Hele-Shaw cell filled with colder, more viscous fluid, shown in blue.}
  \label{fig:model}
\end{figure}

We consider the injection of a hot, less-viscous fluid into a planar channel initially filled with a colder, more viscous fluid. The channel consists of a thin gap between two confining plates that are thermally coupled to a reservoir held at a constant cold temperature. The channel is sketched in three dimensions in figure \ref{fig:model}(\textit{a}). Here, we seek to obtain an effective two-dimensional model, sketched in Figure \ref{fig:model}(\textit{b}).

\subsection{Model description and derivation}
\label{sec:model_description}

We assume that the flow is in the creeping regime \citep{Feder2022}, which means that the Reynolds number $\Reynolds = \rho U h/\mu \ll 1$, where $\rho$ is the fluid density, $U$ is a characteristic fluid velocity, $h$ is the channel half-height, and $\mu$ is the dynamic viscosity of the fluid. The flow is then assumed to be governed by the Stokes equations, leading to the following mass and momentum balance:
\begin{align}
    \div \vec u &= 0, \label{eq:mass_cons_3d}\\
    \div [ 2 \mu (T) \gv \varepsilon ] &= \grad p.  \label{eq:momentum_cons_3d}
\end{align}
Here, $\vec u =\vec u(x,y,z)$ is the velocity field, $p = p(x,y,z)$ the pressure field, the dynamic viscosity $\mu$ is a function of the temperature $T = T(x,y,z)$, while the symmetrized velocity gradient is given by $\gv \varepsilon = (\grad \vec u+\grad \vec u^\top)/2$. 
In this work, we use a first-order approximation of the Arrhenius model for the dynamic viscosity \citep{Blatt2006}, assuming an exponential dependence on temperature:
\begin{equation}
    \ln \left(\frac{\mu(T)}{\mu_c} \right) = \frac{T - T_c}{T_h - T_c}\ln \left(\frac{ \mu_h}{\mu_c}\right),
\label{eq:viscosity}
\end{equation}
where the subscripts $c, h$ refer to the \textit{cold} temperature of the external reservoir and the \textit{hot} temperature of the injected fluid, respectively.
In equation \eqref{eq:momentum_cons_3d} we also neglect buoyancy, that is, we assume that the Rayleigh number $\Rayleigh = \rho \delta g h^3 (T_h - T_c)/(\kappa \mu_c) \ll 1$, where $\delta$ is the thermal expansion coefficient and $g$ is the gravitational acceleration. This assumption implies that thermal convection is negligible compared to thermally induced viscosity-driven flow \citep{Griffiths1986}.

The temperature $T$ evolves in time through the thermal advection-diffusion equation,
\begin{equation}
    \pd{T}{t} + \vec u \bdot \grad T - \kappa \laplacian{T} = 0
\label{eq:advdiff_3d}
\end{equation}
where the fluid's thermal diffusivity $\kappa$ can be obtained from the fluid's heat conductivity, $k_{\mathrm{f}}$, specific heat capacity $c$ and density $\rho$, as $\kappa = k_{\mathrm{f}}/(\rho  c)$. In \eqref{eq:advdiff_3d} we neglect viscous heating.

At the inlet, i.e.\ $x=0$ as indicated in Figure \ref{fig:model}(\textit{a}), we impose that the hot fluid is injected according to a laminar flow profile at a constant flow rate, so the boundary conditions for the velocity and the temperature are given by
\begin{equation}
    \vec u(x=0) = \frac{3}{2}U\left(1 - \left(\frac{z}{h}\right)^2\right)\evec_x, \quad
    T(x=0) = T_h,
    \label{eq:bc_inlet_3d}
\end{equation}
where $U$ is the imposed mean velocity, $z$ is the out-of-plane coordinate, and $\evec_x$ is the unit vector in the $x$ direction, i.e.\ the streamwise direction. Using this velocity profile implicitly assumes that cross-channel viscosity variations may be neglected.

At the outlet $x=L_x$, we impose a constant pressure and no thermal flux:
\begin{equation}
    p(x=L_x) = p_0, \quad
    \evec_x \bdot \grad T|_{x=L_x} = 0,
    \label{eq:bc_outlet_3d}
\end{equation}
At the walls $z=\pm h$, the velocity satisfies no-slip and and the temperature satisfies a Robin (convective) boundary condition to the external reservoir via an \emph{overall} heat transfer coefficient $H_{\mathrm{ov}}$:
\begin{equation}
    \vec u(z = \pm h) = 0, \quad
    \evec_z \bdot k_{\rm f} \grad T|_{\pm h} = H_{\mathrm{ov}}(T_c - T(z = \pm h)),
    \label{eq:bc_walls_3d}
\end{equation}
In general, $H_{\mathrm{ov}}$ depends on the thermal properties of both the fluid and the solid plates. A lumped series-resistance model gives an estimation for $H_{\mathrm{ov}}$:
\begin{equation}
\frac{1}{H_{\mathrm{ov}}} =\frac{1}{H_{\mathrm{f}}} +\frac{\delta_{\mathrm{p}}}{k_{\mathrm{p}}} +\frac{1}{H_{\mathrm{res}}}
\label{eq:Hov_series}
\end{equation}
where $H_{\mathrm{f}}$ is the heat transfer coefficient related to the fluid in the channel, $k_{\mathrm{p}}$ the plate conductivity, $\delta_{\mathrm{p}}$ the thermal penetration depth in the plate, and $H_{\mathrm{res}}$ the external coefficient from the plate’s outer face to the reservoir. For a fully developed laminar flow between parallel plates one may write $H_{\mathrm{f}}=\mathrm{Nu}\,k_{\mathrm{f}}/(2h)$, where the Nusselt number $\mathrm{Nu}$ takes the value $\mathrm{Nu}\approx 7.54$ in case of isothermal walls or $\mathrm{Nu}\approx 8.23$ assuming uniform wall heat flux \citep{Shah2014}. Assuming the external reservoir to be at a constant temperature $T_c$ implies $H_{\mathrm{res}}\to \infty$.

\begin{figure}
    \centering
    \includegraphics[width=0.7\linewidth]{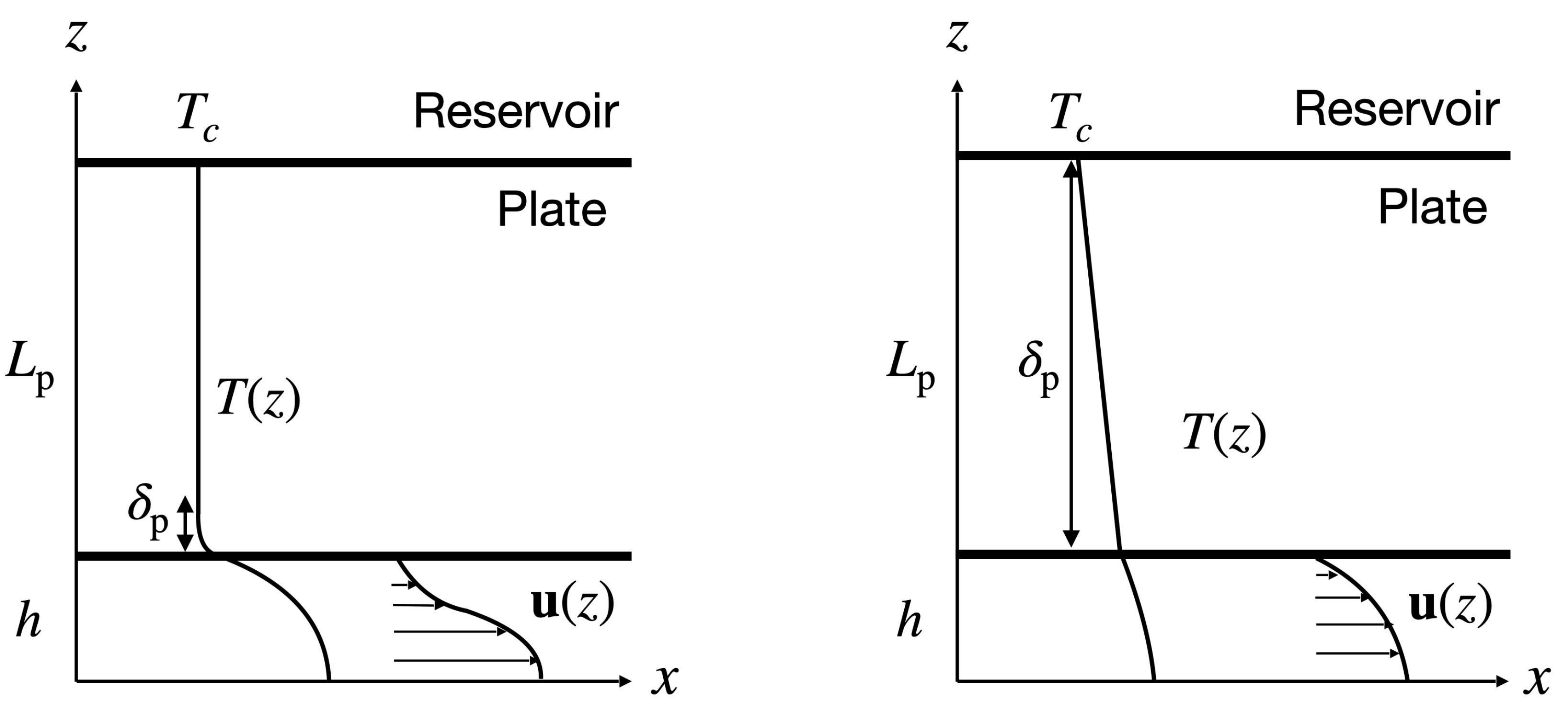}
    \caption{Schematic view of the short-time regime (left) and long-time regime (right), showing temperature profiles $T(z)$ and velocity profiles $\mathbf{u}(z)$ across the plate of thickness $L_{\rm p}$ and the gap of half width $h$, as well as the thermal penetration depth $\delta_{\mathrm p}$ in the plate.
    }
    \label{fig:sketch_cross_section}
\end{figure}

Because the main flow is parallel to the plates, cross-gap heat transport in the fluid is diffusive. Knowing the plates thermal diffusivity $\kappa_{\rm p}$, we can define the fluid cross-gap diffusion time $t_{\rm f}=h^2/\kappa$ and the plate diffusion time $t_{\mathrm{p}}=L_{\mathrm{p}}^2/\kappa_{\rm p}$, and we can identify two limit regimes.
\begin{itemize}
    \item At shorter times $t \sim t_{\rm f}$, the thermal penetration depth in the plate $\delta_{\mathrm{p}}\sim 2\sqrt{\kappa_s t}$ is small compared to $L_{\mathrm{p}}$, so the plate near the interface remains close to $T_c$. The wall may be approximated as isothermal, $T(z=\pm h)\approx T_c$ i.e. a Dirichlet condition. In this limit, the overall thermal resistance is dominated by the fluid side and from \eqref{eq:Hov_series} we get $H_{\mathrm{ov}}\approx H_{\mathrm{f}} \approx k_{\mathrm{f}}/h$. Several previous numerical works on thermo-viscous instabilities have focused on this regime and assumed constant temperature at the fluid-plate interface \citep{Helfrich1995, Wylie1995, Morris1996, Balmforth2004, Holloway2006, Taylor-West2025}.
    \item At longer times $t \gtrsim t_{\rm p} \gg t_{\rm f}$, the plate develops a quasi-steady internal gradient from the interface to the reservoir. If the plate resistance dominates the series, i.e. $L_{\rm p}/k_{\rm p}\gg 1/H_{\mathrm{f}}$, then $H_{\mathrm{ov}}\approx k_{\rm p}/L_{\rm p}.$ In this regime, the plate is not isothermal since it carries a significant temperature drop, while the fluid temperature remains nearly uniform across $z$ (though streamwise variations can persist).
\end{itemize}
In Figure \ref{fig:sketch_cross_section}, we show a sketch of the cross section of the system in the two limit regimes, together with the corresponding temperature profiles and the fluid velocity field in the gap.

We can define the gap Biot number $\Biot$ as the ratio between the fluid cross-gap thermal resistance and the wall resistance
\begin{equation}
\Biot =  \frac{H_{\mathrm{ov}}\,h}{k_f}\propto \frac{1/H_{\mathrm{f}}}{1/H_{\mathrm{ov}}} .
\label{eq:Bi_gap_time}
\end{equation}
At $t \sim t_{\rm f}$ the fluid cross-gap conduction dominates and we have $\Biot \sim 1$, while for $t \gg t_{\rm f}$ the wall resistance dominates, so $\Biot \ll 1$.

To find an effective 2D description of the system, we proceed by averaging the Eq.s \eqref{eq:mass_cons_3d}, \eqref{eq:momentum_cons_3d} and \eqref{eq:advdiff_3d} over the direction perpendicular to the plates, i.e. the $z$-direction. It is then natural to introduce the following gap-averaged quantities indicated by the subscript symbol $\parallel$:
\begin{equation}
    p_{\parallel} = \frac{1}{2 h} \int_{-h}^{h}\! p(x,y,z)\, dz, \quad
    \vec u_{\parallel} = \frac{1}{2 h} \int_{-h}^{h}\! \vec u(x,y,z)\, dz, \quad
    T_{\parallel} = \frac{1}{2 h} \int_{-h}^{h}\! T(x,y,z)\, dz.
    \label{eq:avg_quantities}
\end{equation}
The averaging of the governing equations is detailed in Appendix \ref{sec:averaging}, and leads to the following system
\begin{gather}
    \grad_\parallel \bdot \vec u_\parallel = 0, \label{eq:mass_cons_dim} \\
    -\frac{3}{h^2} \mu(T_\parallel) \vec u_\parallel  = \grad_\parallel p_\parallel, \label{eq:momentum_cons_dim} \\
    \pd{T_\parallel}{t} + \vec u_\parallel \bdot \grad_\parallel T_\parallel -\kappa \laplacian_\parallel T_\parallel -
     \frac{2}{105}\frac{h^2}{\kappa} \grad_\parallel \bdot \left[ \vec u_\parallel \otimes \vec u_\parallel  \grad_\parallel T_\parallel \right] + \frac{H_\mathrm{ov}}{h\rho c}(T_\parallel - T_c)  = 0, \label{eq:advdiff_dim}
\end{gather}
where $\grad_{\parallel} = \evec_x \partial_x + \evec_y \partial_y$ is the gradient operator in the plane.

To simplify the notation and isolate the key parameters influencing the model, we introduce the following non-dimensional variables:
\begin{equation}
   \tilde{\vec x} = \frac{(x,y)}{h},\quad \tilde{\vec u} = \frac{\vec u_\parallel}{U},\quad \tilde{t} = \frac{U}{h} t,\quad 
    \tilde{T} = \frac{T_\parallel - T_c}{T_h - T_c},\quad \tilde{p}= \frac{h}{3\mu_c U}p_\parallel
\label{eq:non-dimensional_var}
\end{equation} 
We also define the viscosity ratio $\beta = \mu_h/\mu_c$, the rescaled cooling rate $\Gamma = H_{\mathrm{ov}} / (\rho c U)$ and the Péclet number $\Peclet = Uh/\kappa$. Leaving out all the tildes for brevity, Equations \eqref{eq:mass_cons_dim}, \eqref{eq:momentum_cons_dim} and \eqref{eq:advdiff_dim} can be expressed in a dimensionless form as
\begin{subequations}
\begin{gather}
    \div \vec{u} = 0, \label{eq:mass_cons} \\
    \vec{u} = -\beta^{-T} \grad p, \label{eq:momentum_cons} \\
    \pd{T}{t} + \vec u \bdot \grad T - \div [(\kappa\, \mathds{I} + \kappa_{\parallel} \, \vec{u} \otimes \vec{u}) \grad T] + \Gamma T = 0, \label{eq:advdiff}
\end{gather}\label{eq:full_model_2d}\end{subequations}
where $\kappa = \Peclet^{-1}$ and $\kappa_{\parallel} = 2 \Peclet / 105$.
The boundary conditions are
\begin{subequations}
\begin{align}
    &\vec u(x=0) = \evec_x, \quad
    T(x=0) = 1, \quad  \text{at the inlet,}
    \label{eq:bc_inlet_2d}
    \\
    &\vec p(x=L_x) = 0, \quad
    \pd{T}{x}\bigg|_{x=L_x} = 1\quad \text{at the outlet.}
    \label{eq:bc_outlet_2d}
\end{align}\label{eq:bc_2d}
\end{subequations}
The effective two-dimensional flow is governed by the Darcy law \eqref{eq:momentum_cons}, where the mobility increases with temperature. Equation \eqref{eq:advdiff} is an advection-dispersion equation for the temperature field, with a local proportional loss term $\Gamma T$, representing the heat conducted through the confining plates, and an in-plane dispersion term which is the sum of two contributions. The first term, which is proportional to $\kappa$, represents the isotropic thermal diffusion, while the second term, which is proportional to $\kappa_\parallel$, results from hydrodynamic (Taylor) dispersion and enhances the effective diffusion only in the direction parallel to the velocity. The Péclet  number controls both these contributions, as $\kappa$ and $\kappa_\parallel$ are directly and inversely proportional to $\Peclet$, respectively. In particular, thermal diffusion is important for $\Peclet\ll 1$, while for $\Peclet\gg 1$, Taylor dispersion dominates the in-plane heat transport. Importantly, we note that in no limit of the  Péclet number does a two-dimensional description remain consistent if the diffusive/dispersive terms are neglected.

We emphasize that the long-time asymptotic limit assumed here requires a small Biot number. In our variables $\Biot=\Gamma\,\Peclet$, hence the gap-averaged model \eqref{eq:full_model_2d} applies only if
\begin{equation}
    \Gamma\,\Peclet \ll 1,
    \label{eq:condition}
\end{equation}
i.e. the wall/reservoir coupling is weak relative to cross-gap conduction in the fluid.

The nonlinearly coupled partial differential equations \eqref{eq:full_model_2d} constitute the full gap-averaged model, in contrast to the linearized model discussed further in \S\,\ref{sec:lsa}.

\subsection{Base state}
\label{sec:model_base_state}
To describe how the Péclet number, viscosity ratio, and heat conduction control the stability of the flow, we first characterize the \emph{base state} of the model, which is what we denote as the steady solution of the model \eqref{eq:full_model_2d} in the absence of any perturbations to the inlet condition \eqref{eq:bc_inlet_2d}.

In the absence of any variations along the in-plane direction $y$ parallel to the inlet, the model is effectively one-dimensional. The steady base state is then obtained solving Eq. \eqref{eq:advdiff} while setting the time derivative to $\partial T/\partial t = 0$ and imposing constant velocity along the $x$-direction. This leads to the following base-state velocity, temperature and pressure, respectively:
\begin{equation}
    \vec u_0(x) = \evec_x, \quad
    T_0(x) = e^{-\xi x}. \quad
    p_0(x) = -\int_0^x \beta^{e^{\xi x'}} dx' + C,
\label{eq:base_state}
\end{equation}
where $C$ is a gauge constant that can be set to $0$. We note in particular that the base-state temperature profile follows a negative exponential with characteristic length $1/\xi$, where
\begin{equation}
\xi = \frac{- 1 + \sqrt{1 + 4\Gamma \kappa_{\rm eff}}}{2\kappa_{\rm eff}}.
\label{eq:def_xi}
\end{equation}
where we define $\kappa_{\rm eff} = \kappa + \kappa_{\parallel} = 1/\Peclet + 2 \Peclet / 105$.
\section{Finger evolution using the full model}
\label{sec:full}
To investigate the response of the system to perturbations, we introduce deviations from the base state and examine how the flow evolves over time. A natural way to implement such perturbations is to modify the inlet boundary conditions \eqref{eq:bc_2d}. The governing equations \eqref{eq:full_model_2d} are then solved numerically in time, allowing us to track the spatial-temporal evolution of the temperature, pressure, and velocity fields. In what follows, we consider two types of perturbation: sinusoidal perturbations that consist of a single mode and random perturbations that consist of a spectrum of modes.
The numerical finite element method used to solve the model in both cases is described in Appendix \ref{sec:num_meth_full}.

\subsection{Sinusoidal perturbation}\label{sec:full_sin}
We first consider instabilities arising from a sinusoidal perturbation of the base state. We prepare the system in the base state \eqref{eq:base_state} at time $t=0$, and then change the inlet velocity boundary condition for a small time interval $t \in (0, t_{\rm pert}]$. The sinusoidal perturbation in the velocity field can then be written as follows:
\begin{equation}
   \vec u(x=0, y,t) = u_x \evec_x,\quad u_x =
    \begin{cases}
        1 + \epsilon \cos(k(y - \pi/k))  & {\rm for }\ t \leq t_{\rm pert}\\
        1 & {\rm for }\ t > t_{\rm pert},\\
    \end{cases}\label{eq:sinusoidal_pert}
\end{equation}
where $\epsilon$ is the perturbation amplitude and $k$ is its wavenumber. We expect the outcome not to depend on $\epsilon$ and $t_{\rm pert}$, as long as we take $\epsilon \ll 1$ and we look at the system for times sufficiently larger than $t_{\rm pert}$.
A simulation is thus fully characterized by a specific combination of $\Peclet$, $\Gamma$, $\beta$ and $k$.

To reduce the influence of lateral boundaries, we apply periodic boundary conditions at the transverse sides of the domain, i.e., at $y=0$ and $y=L_y$ (see Fig. \ref{fig:model}). For a given value of $k$, the transverse system size is set to $L_y=2\pi/k$, ensuring that the perturbation spans exactly one wavelength.

To see if an instability arises, we observe the behavior of the system at times sufficiently larger than $t_{\rm pert}$. For different combinations of the global parameters and the wavenumber, an instability emerges and leads to the formation of a \textit{thermal finger}. The system progressively moves away from the base state until it reaches a stationary state for $t\gg t_{\rm pert}$.

\begin{figure}
    \centering
    \includegraphics[width=1.\linewidth]{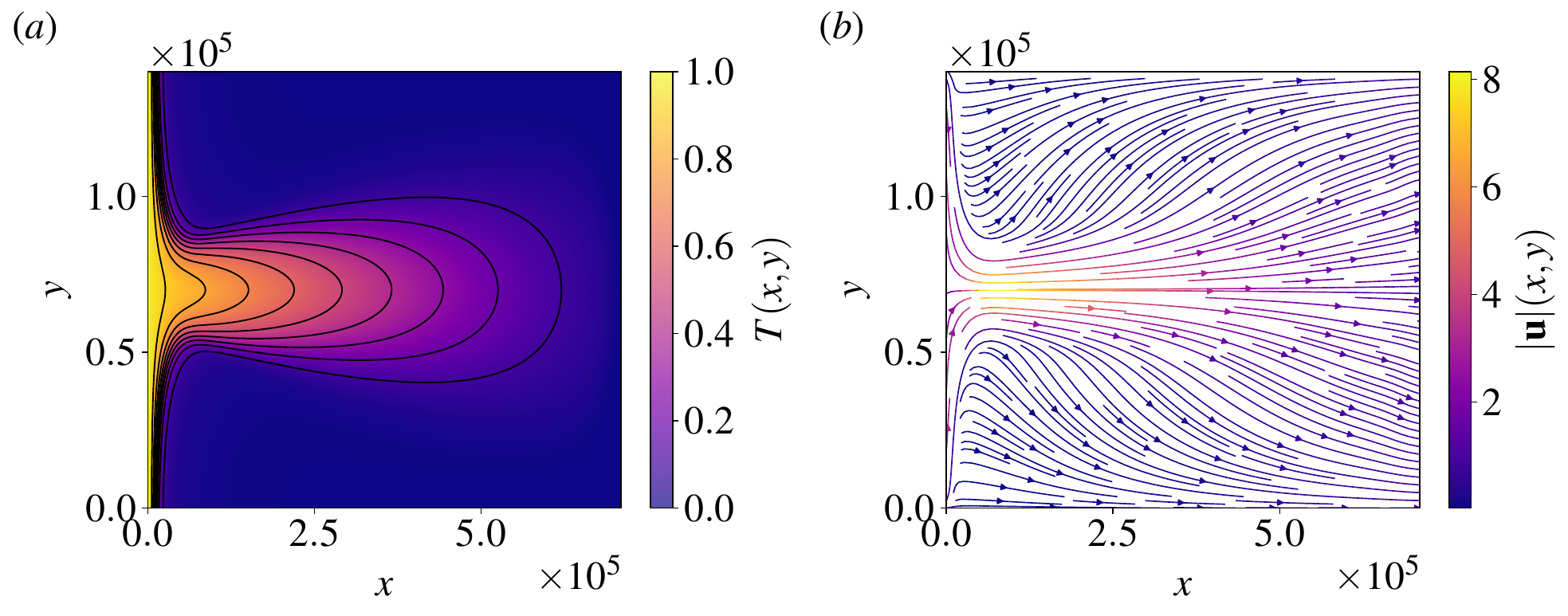}
    \caption{
    Instability emerging from a sinusoidal perturbation. (\textit{a}) Temperature field $T(x,y)$ for $t = 1.2\cdot 10^6 \gg t_{\rm pert}$, where different black contours represent different isothermal curves at $T=0.1,0.2,\dots,0.9$. (\textit{b}) Streamlines of the velocity field $\vec u$, where the local color represents the velocity magnitude $|\vec u|(x,y)$. For this simulation we set $\Peclet = 10^3$, $\Gamma = 10^{-5}$, $\beta = 10^{-3}$ and $k = 2\pi/(1.4\cdot10^{5})$. The perturbation amplitude and time are $\epsilon = 10^{-3}$ and $t_{\mathrm{pert}}=10^3$.}
    \label{fig:Tlevel_and_streamlines_sin}
\end{figure}

An example is shown in Fig. \ref{fig:Tlevel_and_streamlines_sin}, where the temperature and velocity profiles of a system in such a state are shown. For this simulation, we set the global parameters to $\Peclet = 10^3$ and $\Gamma = 10^{-5}$, such that $Pe \Gamma =10^{-2}\ll 1$ following \eqref{eq:condition}, and $\beta = 10^{-3}$. The perturbation wavenumber, amplitude and time are $k = 2\pi/(1.4\cdot10^{5})$, $\epsilon = 10^{-3}$ and $t_{\rm pert}=10^3$. The temperature field $T(x,y)$ is shown in Fig. \ref{fig:Tlevel_and_streamlines_sin}(\textit{a}). The deviation from an exponential profile uniform along $y$ is clearly visible, with a high-temperature region arising close to the inlet at the position of the maximum of the perturbation ($y=7\cdot10^4$), whose contours, emphasized by the isothermal curves, resemble a finger shape. The effects of the instability are also visible by looking at the velocity streamlines in figure \ref{fig:Tlevel_and_streamlines_sin}(\textit{b}), where the flow that is focused in correspondence to the finger region strongly deviates from a uniform flow profile. Far from the inlet, for $x$ sufficiently high, the effects of the perturbation vanish and the temperature and velocity field reach uniformity.

\begin{figure}
    \centering \includegraphics[width=1.\linewidth]{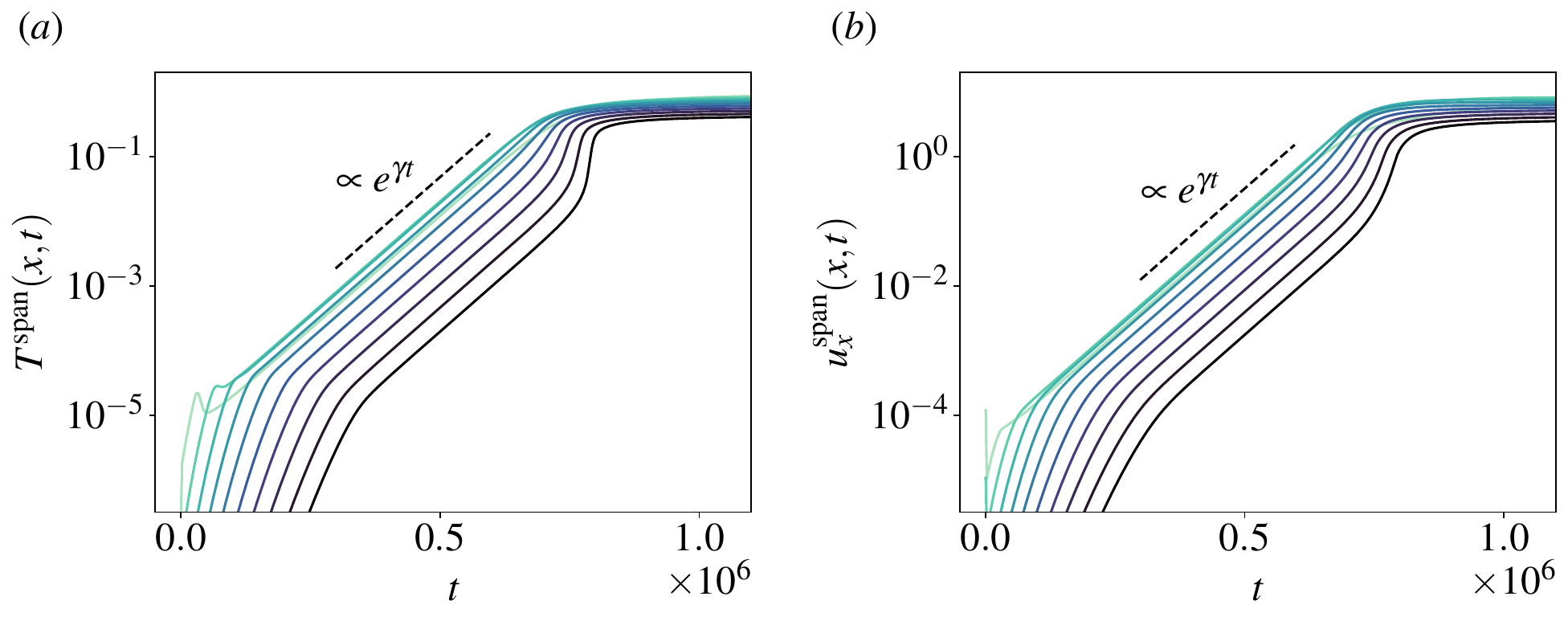}
    \caption{Temporal evolution of the spans in (\textit{a}) temperature, $T^{\rm span}(x,t)$ (see \eqref{eq:Tspan}), and (\textit{b}) streamwise velocity, $u_x^{\rm span}(x,t)$ (see \eqref{eq:uxspan}), for a sinusoidal inlet velocity perturbation. The global parameters and the perturbation wavelength, amplitude and time are the same of Figure \ref{fig:Tlevel_and_streamlines_sin}. Different positions $x$ are represented using a color gradient ranging from light blue ($x=4\cdot10^4$) to dark blue ($x=4\cdot10^5$), with values equally spaced by $4\cdot10^4$. The dashed line indicates the slope of the curves obtained by a best fit procedure in the corresponding time interval. For this simulation, we measured $\gamma \approx 1.61\cdot10^{-5}$.}
    \label{fig:Tspan_and_uxspan}
\end{figure}
To quantify the onset and growth of an instability, an effective way is to consider the difference between the maximum and minimum of a relevant physical quantity. With this aim, we measure the span between the maximum and minimum of both the temperature field and the $x$-component of the velocity field, denoted $T^{\rm span}$ and $u_x^{\rm span}$, respectively: 
\begin{subequations}
\label{eq:T_and_ux_span}
\begin{align}
    T^{\rm span}(x,t) &= T^{\max}\left(x,t\right) - T^{\min}\left(x,t\right),\label{eq:Tspan}\\
    u_x^{\rm span}(x,t) &= u_x^{\max}\left(x,t\right) - u_x^{\min}\left(x,t\right).\label{eq:uxspan}
\end{align}\end{subequations}
Here, the superscripts $\max$ and $\min$ refer, respectively, to the maximum and minimum value of the related quantity along $y$, given $x$ and $t$. For the sinusoidal perturbations considered in this subsection, we expect the maximum and minimum to occur at $y=\pi/k$ and $y=0$ (or $2\pi/k$), respectively, for all $x$ and $t$.

Initially and as long as the relative amplitude of the perturbation remains small, we expect the instability to evolve exponentially in time, i.e.\ like
\begin{equation}
T^{\rm span}(x,t) \propto e^{\gamma t}, \quad u_x^{\rm span}(x,t) \propto e^{\gamma t}
\end{equation}
where $\gamma$ is the growth rate of the instability.  If $\gamma < 0$, the perturbation decays, meaning that the base state is stable, while if $\gamma > 0$ the system is unstable and the exponential behavior holds as long as non-linear growth-limiting effects remain negligible.

Figure \ref{fig:Tspan_and_uxspan} shows the temporal evolution of both the temperature and velocity spans for an unstable system.
We see that, after a short transient phase, both quantities grow exponentially in time, corresponding to the linear growth phase, until they slow down and eventually saturate at a plateau corresponding to a stationary state distinct from the base state.

The exponential behavior holds for different choices of $x$ sufficiently close to the inlet. From a linear fit of the couples $(\log T^{\rm span},\ t)$  (or $(\log u_x^{\rm span},\ t)$), in a time interval corresponding to the linear phase, it is then possible to measure the growth rate $\gamma$.

\subsection{Random perturbation}\label{sec:full_random}
In real physical systems, perturbations are rarely limited to a single mode, but rather consist of a superposition of several modes with random phases and amplitudes. To mimic natural perturbations, we now consider a white-noise perturbation in which all spatial modes are present with equal amplitude but random phase. This initial condition allows all modes to evolve simultaneously, with the unstable ones growing and eventually dominating the behavior of the system.

We implement a white-noise perturbation by the following boundary condition for the velocity at the inlet:
\begin{equation}
    \vec u(x=0,y,t) =  u_x\vec e_x, \quad u_x = \begin{cases}
1 + \epsilon \eta(y)   & {\rm for }\ t \leq t_{\rm pert}\\
1 & {\rm for }\ t > t_{\rm pert}\\
\end{cases}\label{eq:random_pert}
\end{equation}
where $\eta(y)$ is an uncorrelated random variable distributed according to a Gaussian distribution centered at $\mean{\eta}=0$ and with a variance $\mean{\eta^2}=1$. Like in \S\,\ref{sec:full_sin}, we implement periodic boundary conditions at $y=0$ and $y=L_y$, although in this case the transverse system size $L_y$ is not related to any specific perturbation wavelength.

\begin{figure}
    \centering
    \includegraphics[width=1.\linewidth]{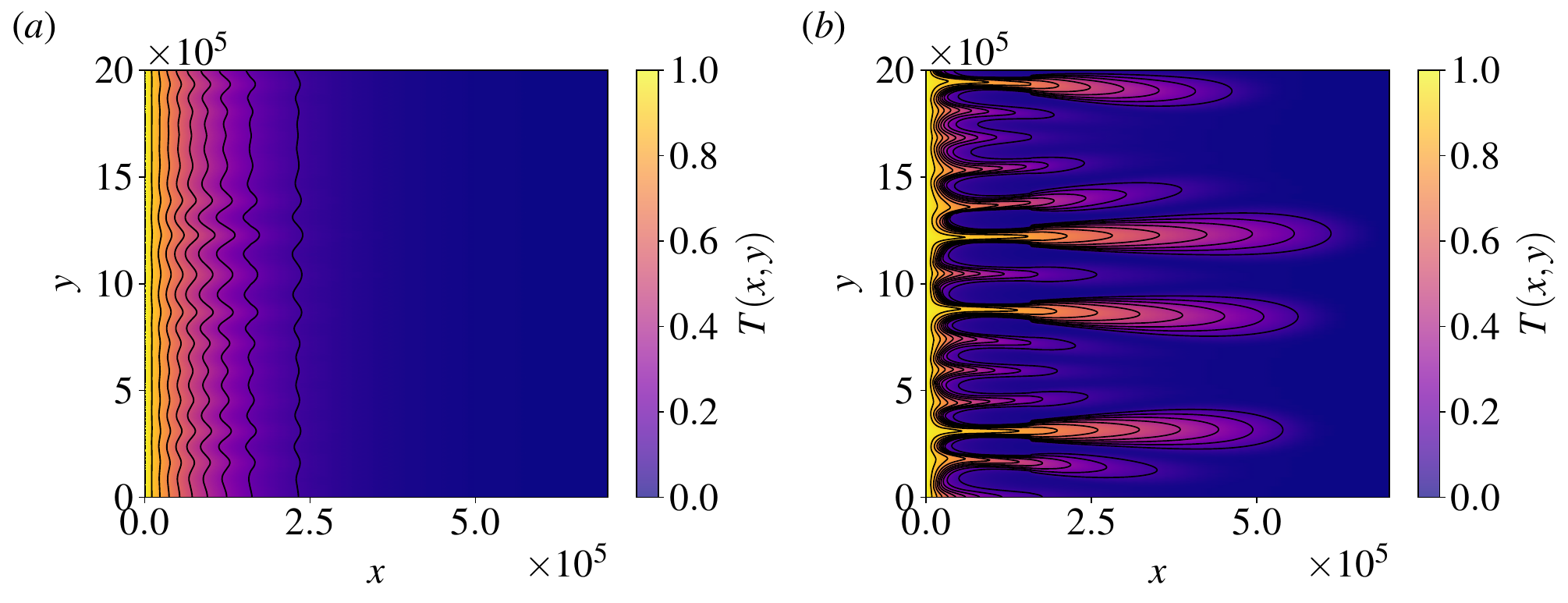}
    \caption{Instability emerging from a random perturbation. The temperature field $T(x,y, t)$ is shown at two time instances (\textit{a}) $t = 1.375\cdot 10^6$ and (\textit{b}) $t = 1.75\cdot 10^6$. The black contours represent different isothermal curves at $T=0.1,0.2,\dots,0.9$. (\textit{b}). For this simulation we set $\Peclet = 10^3$, $\Gamma = 10^{-5}$, and $\beta = 10^{-3}$. The perturbation amplitude and time are $\epsilon = 10^{-3}$ and $t_{\mathrm{pert}}=10^3$.}
    \label{fig:Tlevels_rnd}
\end{figure}
\begin{figure}
    \centering
    \includegraphics[width=1.\linewidth]{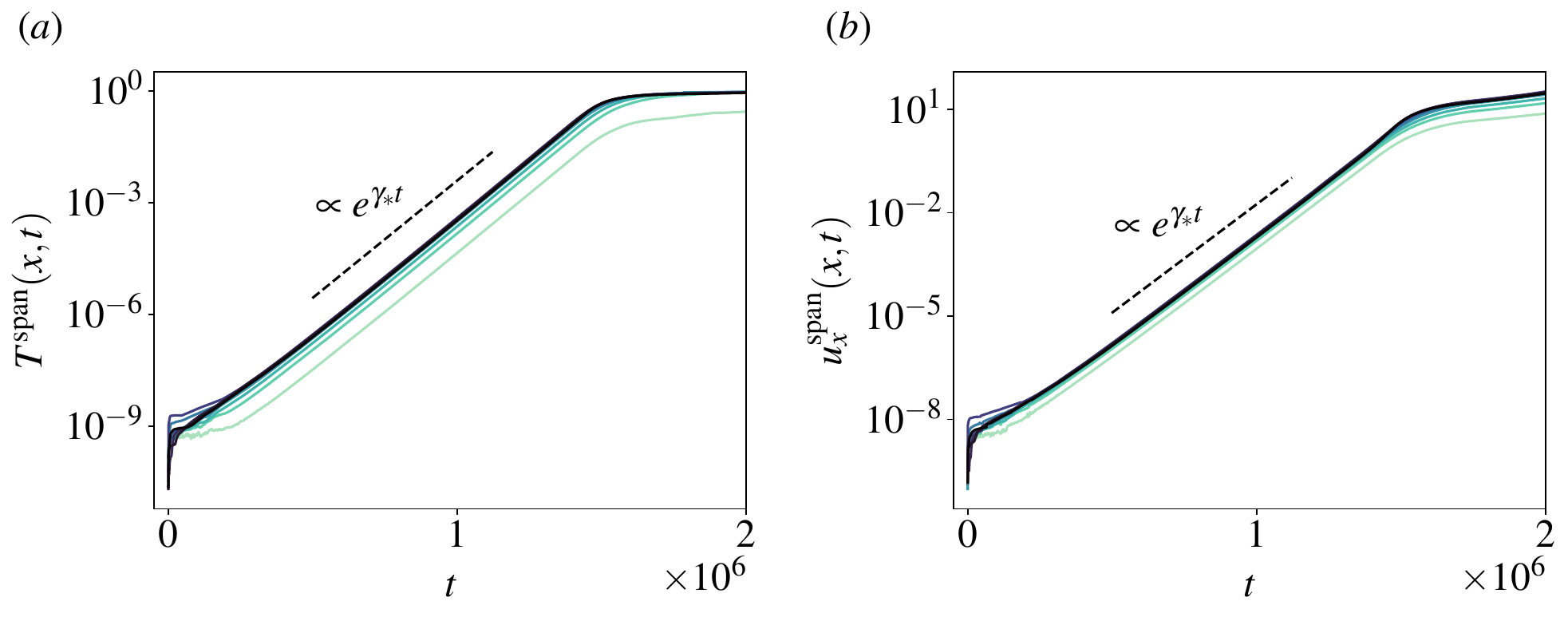}
    \caption{Temporal evolution of the span in (\textit{a}) temperature, $T^{\rm span}(x,t)$ (see \eqref{eq:Tspan}), and (\textit{b}) streamwise velocity, $u_x^{\rm span}(x,t)$ (see \eqref{eq:uxspan}), for a random inlet velocity perturbation. The global parameters and the perturbation amplitude and time are the same of Figure \ref{fig:Tlevels_rnd}. Different positions $x$ are represented using a color gradient ranging from light blue ($x=4\cdot10^4$) to dark blue ($x=4\cdot10^5$), with values equally spaced by 1. The dashed line indicates the slope of the curves obtained by a best fit procedure in the corresponding time interval.}
    \label{fig:Tspan_and_uxspan_rnd}
\end{figure}

Figure \ref{fig:Tlevels_rnd} shows the development of the instability triggered by an initial random perturbation, in a single simulation with parameters $\Peclet = 10^3$, $\Gamma = 10^{-5}$, and $\beta = 10^{-3}$. The panels (\textit{a}) and (\textit{b}) show the temperature field $T(x,y)$ at times $t = 1.375\cdot 10^6$ and $t = 1.75\cdot 10^6$, both significantly larger than the perturbation time $t_{\mathrm{pert}}=10^3$. The evolving distortion of the temperature field is emphasized by the isothermal curves.

In panel (\textit{a}), corresponding to the linear growth regime, only a subset of modes is visibly amplified. This selection becomes even clearer in panel (\textit{b}), which represents a later time when nonlinear effects have become significant: The amplitude of the most unstable mode has increased substantially, and its structure dominates the pattern. The dominant wavelength $k_*$ is estimated from the number of emerging fingers $n_f$ visible in the interval $0<y<L_y$, using the relation $k_* = 2 \pi n_f/L_y$.

To quantify the growth of the instability, we track the evolution of the spans $T^{\mathrm{span}}(x,t)$ and $u_x^{\mathrm{span}}(x,t)$, defined in equation \eqref{eq:T_and_ux_span}. Figure \ref{fig:Tspan_and_uxspan_rnd} displays their time evolution for various locations along $x$. Analogously to the sinusoidal perturbation case, a period of exponential growth is evident, indicated by the dotted reference line, between an initial transient period and a final non-linear regime. We find that during the linear growth phase both quantities scale as $e^{\gamma_* t}$, where $\gamma_*$ denotes the growth rate of the fastest mode.

At later times, the stabilization of fingers to steady structures of finite length reflects the balance between the temperature advection inside the cell and the cooling through the walls, and has been addressed in previous numerical studies \citep{Helfrich1995, Diniega2013}. In this paper, we focus on the early stage of the instability and on the physics that controls its emergence in thermo-viscous flow, leaving the characterization of the late-stage steady state a topic for further study.

\section{Linear stability analysis}
\label{sec:lsa}
To investigate the mechanisms underlying the instability observed in the simulations of the full model, we now turn to a linear stability analysis of the governing equations. By considering small perturbations around the base state, the equation system can be linearized, allowing us to examine the growth rates and structure of individual modes in the early, exponential phase of the instability. This approach yields a dispersion relation, from which the maximum growth rate and its associated most unstable wavenumber can be determined.

\subsection{Derivation}
To derive the linearized problem from the full model \eqref{eq:full_model_2d}, we follow a standard procedure (see e.g.\ \citep{Helfrich1995}). We introduce a perturbation around the base state, such that a solution can be written as
\begin{subequations}
    \begin{align}
    p(x) &= p_0(x) + p'(x,y,t) \\
    T(x) &= T_0(x) + T'(x,y,t) \\
    \vec u(x) &= \vec u_0(x) + \vec u'(x,y,t)
\end{align}
\label{eq:perturbation}
\end{subequations}
where $T'$, $\vec u'=(u_x',u_y')$ and $p'$ are the perturbation terms for the temperature, velocity, and pressure, respectively, and we have $T'\sim \epsilon$, $|\vec u'|\sim \epsilon$ and $p'\sim \epsilon$, where $\epsilon \ll 1$ is a small perturbation parameter related to the deviation from the base state at the inlet; see equations \eqref{eq:sinusoidal_pert} and \eqref{eq:random_pert}. Keeping only terms of first order in $\epsilon$, equations \eqref{eq:mass_cons}, \eqref{eq:momentum_cons}, \eqref{eq:advdiff} become:
\begin{subequations}
\label{eq:lsa_prime}
\begin{align}
    \partial^2_x u'_x - \psi \partial_x T_0 \partial_x u'_x + \partial^2_y u'_x &= \psi \partial^2_y T' \label{eq:lsa_prime_darcy} \\
    \partial_t T' + u'_x\partial_xT_0 + \partial_x T' - \kappa^2 \laplacian{T'} - \kappa_{\parallel} (\partial^2_x T' + \partial_x u'_x \partial_x T_0 + 2 u'_x \partial^2_x T_0) &= - \Gamma T'  \label{eq:lsa_prime_advdiff}
\end{align}
\end{subequations}
Here we have defined $\psi = - \log \beta > 0$ for brevity. Equation \eqref{eq:lsa_prime_darcy} is obtained by combining the incompressibility condition $\div \vec u' = 0$ with the perturbed Darcy law, obtained from Eq. \eqref{eq:mass_cons} and Eq. \eqref{eq:momentum_cons} respectively. Using the expression for the base state from Eq. \eqref{eq:base_state}, the perturbed velocity can be expressed as a function of the pressure and temperature perturbation terms:
\begin{align}
    \vec u' = - \beta^{-T_0} \grad p' + \psi T' \hat{\vec x}.
    \label{eq:uprime}
\end{align}
Equation \eqref{eq:lsa_prime_advdiff} is obtained keeping the linear terms of the advection-diffusion equation \eqref{eq:advdiff}. Note that both \eqref{eq:lsa_prime_darcy} and \eqref{eq:lsa_prime_advdiff} present only two independent perturbation fields, namely $T'$ and $u_x'$. A linear instability can then be identified by finding a solution for these two fields. The boundary conditions for \eqref{eq:lsa_prime} are

\begin{equation}
    T'(0, y, t) = T'(\infty, y, t) = T'(x, y, 0) = 0, \quad \text{and}\quad
    u_x'(0, y, t) = \epsilon e^{iky}.\label{eq:bc_T'_u'}
\end{equation}

We now take the Fourier transform in the $y$ direction, i.e.\ express
\begin{align}
    T'(x, y, t) = T_k (x, t) e^{iky}, \quad u_x'(x, y, t) = u_k (x, t) e^{iky},
\end{align}
which inserted into \eqref{eq:lsa_prime} gives an equation system for $T_k$ and $u_k$:
\begin{subequations}
\begin{align}
  \partial_x^2 u_k - \psi \partial_x T_0 \partial_x u_k - k^2 u_k &= - k^2 \psi T_k
    \label{eq:lsa_k_darcy} \\
    \pdt T_k + \partial_x T_k - \kappa_{\rm eff} \partial_x^2 T_k + \left[ \kappa k^2 + \Gamma \right]  T_k
    &= \partial_x T_0 \left( - u_k + \kappa_\parallel \partial_x u_k \right) + 2 \kappa_\parallel u_k \partial_x^2 T_0 
    \label{eq:lsa_k_advdiff}
\end{align}
\label{eq:lsa_k}
\end{subequations}
with the boundary conditions
\begin{subequations}
\begin{gather}
    T_k (0, t) = T_k(\infty, t) = T_k(x, 0) = 0, \\
    u_k(0, y, t) = \epsilon.
\end{gather}\label{eq:bc_Tk_uk}\end{subequations}
The problem \eqref{eq:lsa_k}-\eqref{eq:bc_Tk_uk} is solved numerically using a finite element method (see Appendix \ref{sec:num_meth_lin}).

\begin{figure}
    \centering
    \includegraphics[width=1.\linewidth]{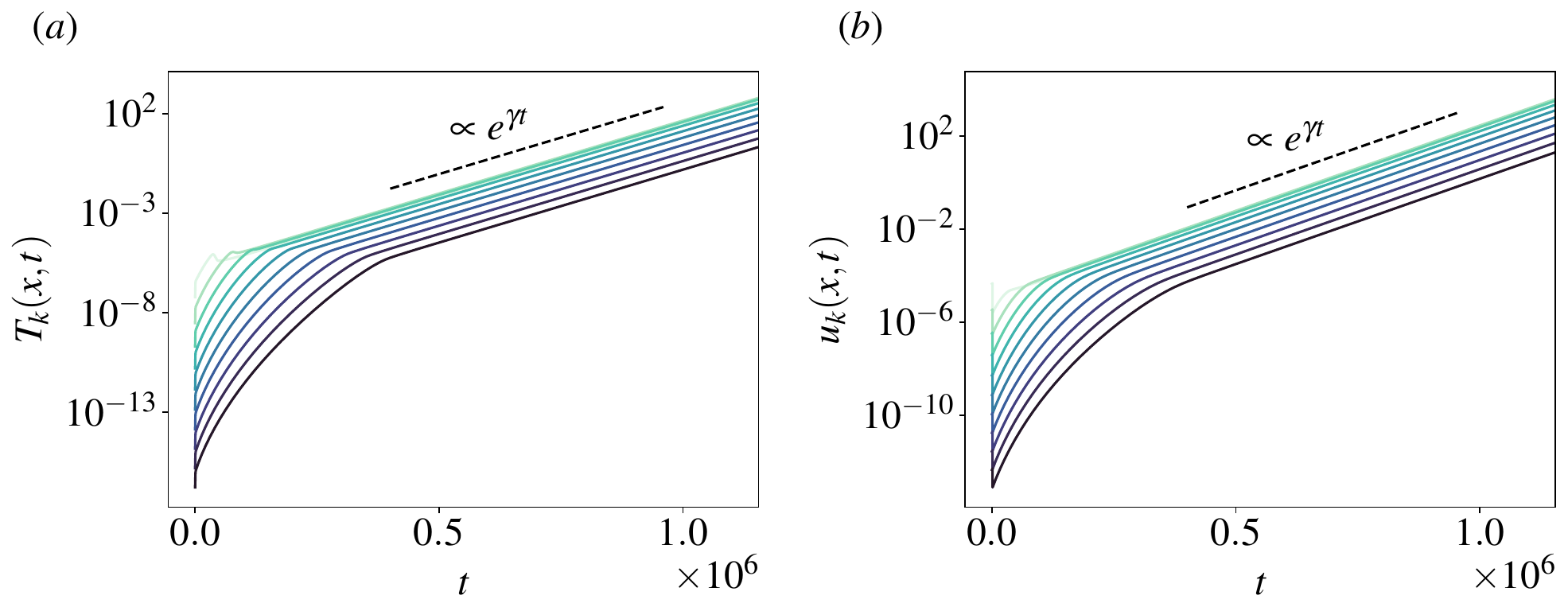}
    \caption{
    Temporal evolution of the Fourier-transformed perturbation term for (\textit{a}) the temperature $T_k$ and (\textit{b}) the streamwise velocity $u_k$. For this simulation, we set $\Peclet = 10^3$, $\Gamma = 10^{-5}$, $\beta = 10^{-3}$ and $k = 2\pi/(1.4\cdot10^{5})$. Different positions $x$ are represented using a color gradient ranging from light blue ($x=4\cdot10^4$) to dark blue ($x=4\cdot10^5$), with values equally spaced by $4\cdot10^4$. The dashed line indicates the slope of the curves obtained by a best fit procedure in the corresponding time interval, from which we measured $\gamma \approx 1.69\cdot 10^{-5}$. The perturbation amplitude and time are $\epsilon = 10^{-3}$ and $t_{\mathrm{pert}}=10^3$.}
    \label{fig:T_and_u_vs_t}
\end{figure}
The solutions for $T_k(x, t)$ and $u_k(x, t)$ are shown as functions of time $t$ for different streamwise positions $x$ in figure \ref{fig:T_and_u_vs_t}, where we have set $\Peclet = 100$, $\Gamma = 1$ and $\beta = 10^{-3}$. After an initial transient interval, both functions follow a steady exponential growth in time. The slope of each curve does not depend on the choice of $x$, indicating that the whole function approaches a constant spatial shape and only its scale evolves in time.

For a given set of parameters, the growth rate $\gamma = \gamma(k)$ related to the spatial frequency $k$ of the perturbation is then measured by a least-squares linear fit to this perturbation scale. For an unbiased measure of the perturbation scale, we use the maximum value of $T_k$ along $x$, i.e.\ $T_{\rm max} (t) = \max_{x} T_k(x, t)$.

Assuming a time-invariant shape and exponential growth of the perturbation is equivalent to writing:
\begin{align}
    T_k(x, t) = \hat T_k (x) e^{\gamma t}, \quad u_k = \hat u_k (x) e^{\gamma t}.
    \label{eq:lsa_hatk_def}
\end{align}
Inserting \eqref{eq:lsa_hatk_def} into equations \eqref{eq:lsa_k} leads to a system of ordinary differential equations (ODE) for $\hat T_k(x)$ and $\hat u_k(x)$:
\begin{subequations}
\begin{align}
    \left( \dddx - \psi \d{T_0}{x} \d{}{x} - k^2\right) \hat u_k  &=  - k^2 \psi \hat T_k, \label{eq:darcy_laplace} \\
    \left(\gamma + \ddx - \kappa_{\rm eff} \dddx + \kappa k^2 + \Gamma \right) \hat T_k &= \left( -\d{T_0}{x} + \kappa_{\parallel} \left(2 \dd{T_0}{x} + \d{T_0}{x} \ddx \right) \right) \hat u_k \label{eq:advdiff_laplace}.
\end{align}\label{eq:lsa_hatk}\end{subequations}
The system \eqref{eq:lsa_hatk} is generally not solvable analytically.
However, in the outer region ($x \gg \xi^{-1}$) we can set $T_0 = 0$, so equations \eqref{eq:darcy_laplace} and \eqref{eq:advdiff_laplace} are simplified as, respectively,
\begin{subequations}
\begin{align}
    \left(\dddx - k^2\right) \hat u_k  &= - k^2 \psi \hat T_k,\\
    \left(\gamma + \ddx - \kappa_{\rm eff} \dddx + \kappa k^2 + \Gamma \right) \hat T_k &= 0.
\end{align}
\label{eq:lsa_hatk_bs0}
\end{subequations}
Applying the boundary condition $\hat T_k(x \to \infty) = 0$, the solution can be written as
\begin{subequations}
\begin{align}
    \hat T_k(x) &= A e^{-\Lambda x}, \label{eq:lsa_hatk_bs0_Tk} \\
    \hat u_k(x) &= \frac{\psi k^2}{k^2 - \Lambda^2} A e^{-\Lambda x} + B e^{-k x} ,\label{eq:lsa_hatk_bs0_uk}
\end{align}\label{eq:lsa_hatk_bs0_sol}\end{subequations}
where $A$ and $B$ are undefined constants and
\begin{equation}
\Lambda = \frac{- 1 + \sqrt{1 + 4(\Gamma + \kappa k^2 + \gamma) \kappa_{\rm eff}}}{2\kappa_{\rm eff}}
\label{eq:Lambda}
\end{equation}
is the inverse of the exponential decay length.
\begin{figure}
    \centering
    \includegraphics[width=1\linewidth]{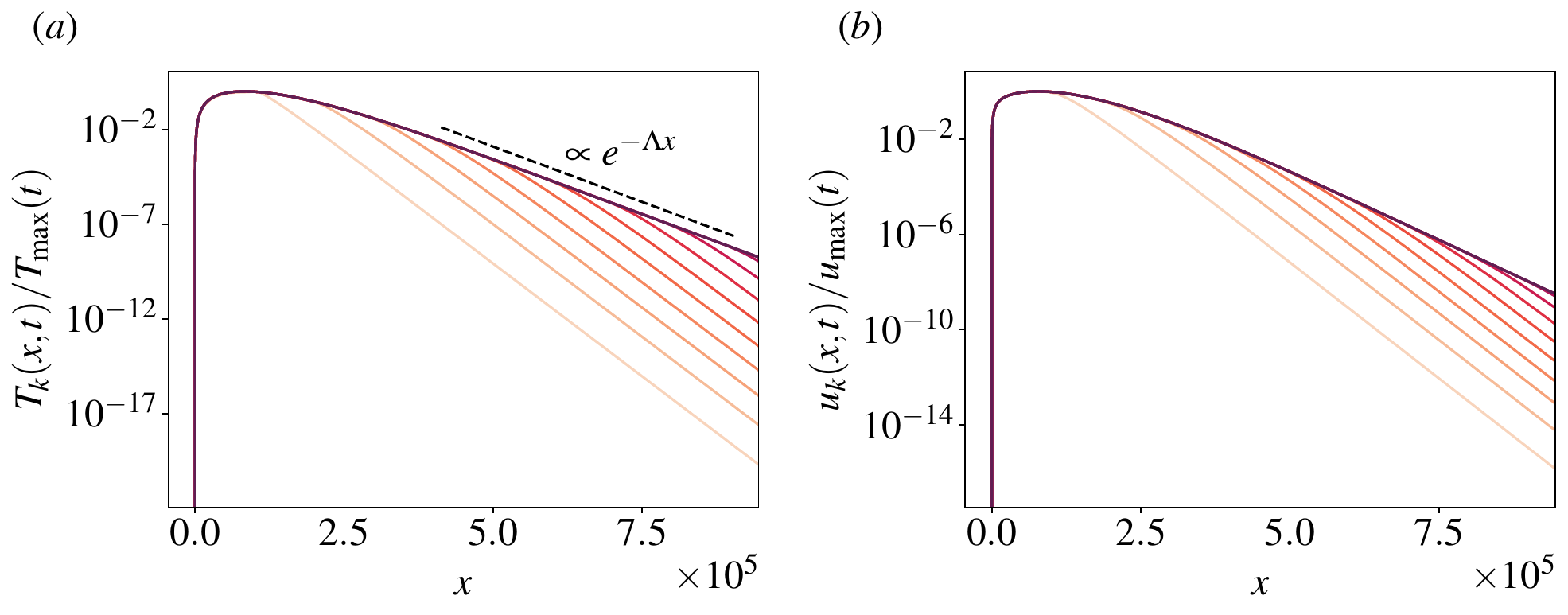}
    \caption{
    Fourier-transformed perturbation term for (\textit{a}) the temperature $T_k$ and (\textit{b}) the streamwise the velocity $u_k$, as a function of the position $x$. The global parameters and the perturbation wavelength, amplitude and time correspond to the ones of Figure \ref{fig:T_and_u_vs_t}. Different times $t$ are represented using a color gradient ranging from light red ($t=10^5$) to dark red ($t=10^6$), with values equally spaced by $10^5$. The dashed line in ($a$) represents the slope of the exponential solution \eqref{eq:lsa_hatk_bs0_Tk} with $\Lambda = 2.690\cdot 10^{-5}$.}
    \label{fig:T_and_u_vs_x}
\end{figure}

In Fig. \ref{fig:T_and_u_vs_x}, we show the evolution along $x$ of the rescaled perturbation terms $T_k(x,t)/T_{\max}(t)$ and $u_k(x,t)/u_{\max}(t)$, where $T_{\max}(t) = \max_x T_k(x, t)$ and $u_{\max}(t) = \max_x u_k(x, t)$.
Both functions increase sharply for small values of $x$, followed by a monotonic decrease that is exponential for large $x$. In particular, the slope of the exponential tail for the temperature term $T_k$ matches, for sufficiently high $t$, the one from the analytical solution \eqref{eq:lsa_hatk_bs0_Tk} valid for $x \gg \xi^{-1}$.

\subsection{Dispersion relations}
\label{sec:lsa_disp_rels}
For any combination of viscosity ratio $\beta$, cooling rate $\Gamma$, Péclet number $\Peclet$, and wavenumber $k$ of the perturbation, it is possible to numerically measure the growth rate $\gamma$ and thereby establish a dispersion relation $\gamma(k; \Peclet, \Gamma, \beta)$.

\begin{figure}
    \centering
    \includegraphics[width=1.0\linewidth]{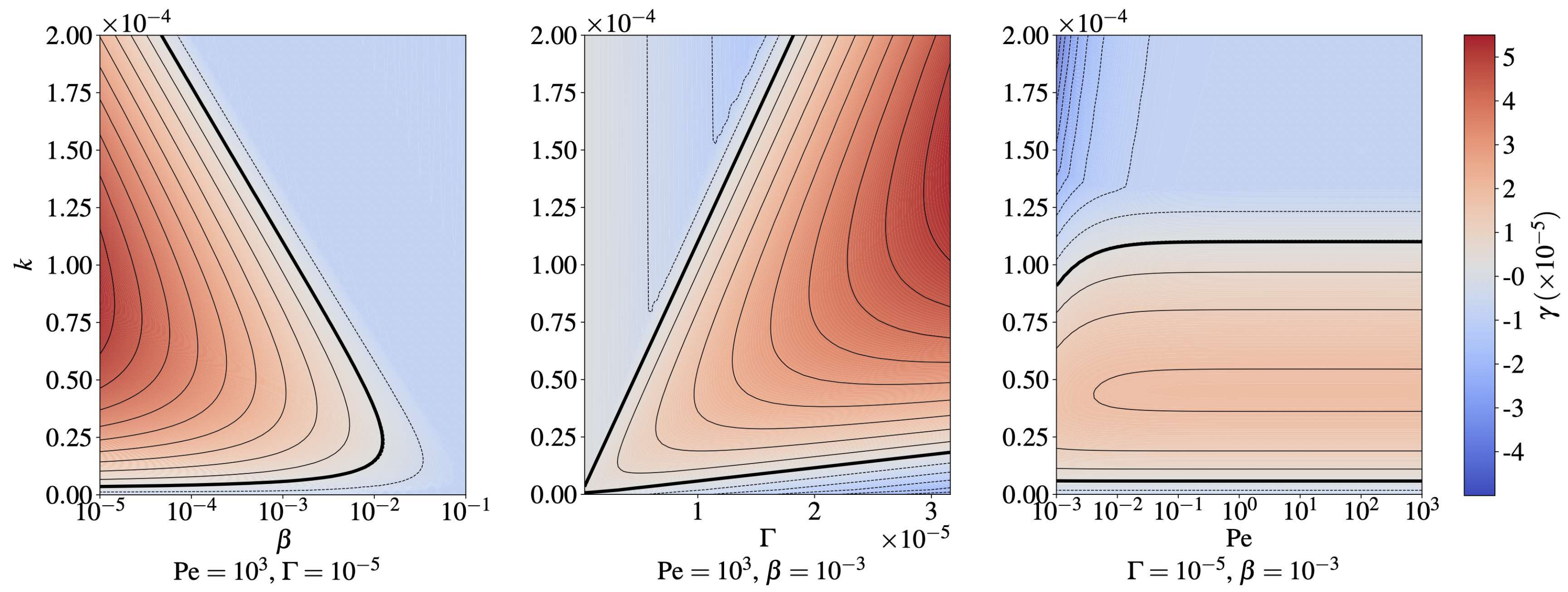}
    \caption{Parameter space of the instability growth rate $\gamma$ as a function of the wavenumber $k$ and the system parameters $\Peclet$, $\Gamma$ and $\beta$. Each panel shows $\gamma(k; \Peclet, \Gamma, \beta)$ for different values of one parameter while keeping the other two fixed: (\textit{a}) viscosity ratio $\beta$, (\textit{b}) cooling rate $\Gamma$, (\textit{c}) Péclet number $\Peclet$. The color indicates the value of $\gamma$, where red shows positive growth and blue shows stable regions. Contours are spaced by $5\cdot 10^{-6}$, and the thicker contour marks the $\gamma = 0$ boundary between stable and unstable regions.}
    \label{fig:gamma_vs_k_linear_cmap}
\end{figure}

Figure \ref{fig:gamma_vs_k_linear_cmap} shows an intensity map of the growth rate $\gamma$ as a function of $\beta$, $\Gamma$, and $\Peclet$ in panels (\textit{a-c}), respectively. For each plotted parameter, the other two are kept constant and fixed to the values shown below. For a given combination of $\Peclet$, $\Gamma$ and $\beta$, the dispersion relation $\gamma(k)$ corresponds to a vertical section of the intensity map based on these values.

The plots in Figure \ref{fig:gamma_vs_k_linear_cmap} reveal a non-trivial dependence of the growth rate on the wavenumber and the three global parameters. For each parameter, the shape of the contours, representing points with the same value of $\gamma$, depends in general on the values of the other two parameters that are held fixed. In the following Sections we will present predictions for the dispersion relations and of their maxima in significant limits. Here, we just make a few preliminary observations.

First, for some combinations of $\Peclet$, $\Gamma$ and $\beta$, $\gamma(k) < 0$ for all $k$, while for other choices of these global parameters, there exists an interval of values of $k$ for which $\gamma(k) > 0$. This implies that for combinations of $\Peclet$, $\Gamma$ and $\beta$ for which $\gamma$ is always negative, an instability cannot develop from any kind of perturbation, and the system is unconditionally stable. On the other hand, if an interval of positive $\gamma$ exists, an instability is expected to grow only under perturbations that contain modes of the corresponding wavenumber $k$. In general, the interval of positive $\gamma$ changes with the combination of $\Peclet$, $\Gamma$ and $\beta$. We note that, for all combinations explored, this interval always present a long-wave and a short-wave cutoff, namely $\gamma < 0$ for sufficiently high, or sufficiently low, $k$.

Second, for all the combinations of parameters studied, cross sections of $\gamma$ along $k$ exhibit exactly one maximum. We denote this maximum growth rate $\gamma_{\max} = \max_k \gamma(k)$, and its corresponding wavenumber $k_{\rm max}$, that is the value of $k$ for which $\gamma(k_{\max}) = \gamma_{\max}$. For a perturbation composed of a spectrum of wavenumbers, we expect the instability to grow at the wavenumber corresponding to the fastest growth rate in the spectrum, at least until non-linear terms start to be relevant. In particular, if the perturbation consists of white noise, as discussed in \S\,\ref{sec:full_random}, an instability will develop at a wavelength equal to $2\pi/k_{\max}$ and will grow exponentially at a rate equal to $\gamma_{\max}$, at least at times sufficiently small for the linear stability approximation to hold. This growth requires that $\gamma_{\max} > 0$; otherwise it will decay. To predict the evolution of instabilities under such conditions, we need to study how $k_{\max}$ and $\gamma_{\max}$ depend on the global parameters.

\subsection{Evolution of the maximum growth rate and corresponding wavenumber}
\label{sec:lsa_gammamax_kmax}
In the general case, we expect that the maximal growth rate $\gamma_{\max}$ and the most unstable wavenumber $k_{\max}$ reflect some of the complexity of the dispersion relation described in the previous Section. However, in the limit of high Péclet number, $\Peclet \gg 1$, we find the analytical predictions (see Appendix \ref{sec:app:largePe}):
\begin{subequations}
\begin{align}
    \gamma_{\rm max} ( \Peclet, \Gamma, \beta ) &\simeq  \Gamma \tilde \gamma_{\rm max} (\beta), \label{eq:gammamax_largePe} \\
    k_{\rm max} ( \Peclet, \Gamma, \beta ) &\simeq \Gamma \tilde k_{\rm max}(\beta), \label{eq:kmax_largePe}
\end{align}\label{eq:gammamax_and_kmax_largePe}\end{subequations}
where $\tilde \gamma_{\rm max}, \tilde k_{\rm max}$ are dimensionless functions of the viscosity ratio alone. For a given $\beta$, both $\gamma_{\rm max}$ and $k_{\rm max}$ are expected to reach a constant value proportional to $\Gamma$ for large $\Peclet$.

In the additional limit of low viscosity ratio, $\beta\ll 1$, we find
\begin{subequations}
\begin{align}
    \tilde \gamma_{\rm max} ( \log \beta ) &\simeq a_\gamma\log\beta + b_\gamma, \label{eq:gammamax_largePe_largepsi} \\
    \tilde k_{\rm max} ( \log \beta ) &\simeq a_k\log\beta + b_k, \label{eq:kmax_largePe_largepsi}
\end{align}\label{eq:gammamax_and_kmax_largePe_largepsi}\end{subequations}
where $a_\gamma$, $b_\gamma$, $a_k$ and $b_k$ are numerical constants (approximate values are given in Table \ref{tab:coefficients_from_fit}). For sufficiently large $\Peclet$, we find these polynomial expressions to yield good agreement with numerical data for all unstable configurations, as we will show in the following. The derivation of \eqref{eq:gammamax_and_kmax_largePe} and \eqref{eq:gammamax_and_kmax_largePe_largepsi} is based on scaling equations \eqref{eq:lsa_hatk}, neglecting small terms, and grouping the independent parameters, as detailed in Appendix \ref{sec:app:largePe}.

We numerically compute how the fastest growth rate and the corresponding wavenumber depends on the global parameters and eventually compare it with \eqref{eq:gammamax_and_kmax_largePe_largepsi}. To find $\gamma_{\max}$ and $k_{\max}$ for a given combination of global parameters $\Peclet$, $\Gamma$, and $\beta$, we compute $\gamma$ by solving \eqref{eq:lsa_k} and \eqref{eq:bc_Tk_uk} for an interval of values of $k$ that includes the maximum, and locate the maximum using a quadratic fit to these values. The coordinate of the maximum from the best fit corresponds to ($k_{\max}$, $\gamma_{\max}$). Next, we discuss how the stability of thermo-viscous flow evolves with the three control parameters $\beta$, $\Gamma$ and $\Peclet$.

\begin{figure}
    \centering
    \includegraphics[width=1.\linewidth]{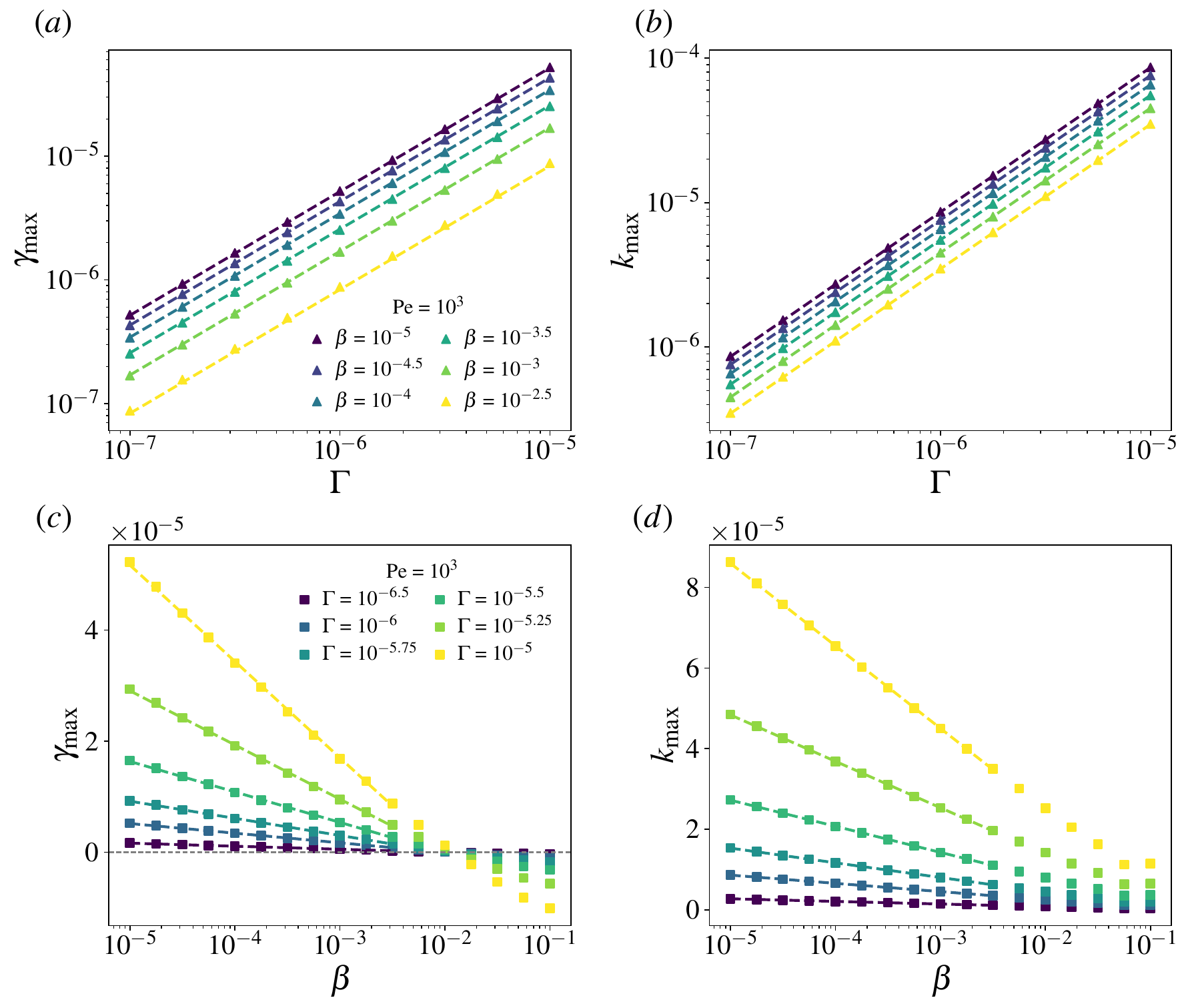}
    \caption{Maximum growth rate $\gamma_{\rm max}$ (panels (\textit{a}), (\textit{c})) and corresponding wavenumber $k_{\rm max}$ (panels (\textit{b}), (\textit{d})) as functions of the viscosity ratio $\beta$ (panels (\textit{a}), (\textit{b})) and of the cooling rate $\Gamma$ (panels (\textit{c}), (\textit{d})), for $\Peclet = 10^3$.
    Panels (\textit{a}) and (\textit{b}) show variations with $\beta \in [10^{-5}, 10^{-1}]$ at fixed $\Gamma$, while panels (\textit{c}) and (\textit{d}) show variations with $\Gamma \in [10^{-7}, 10^{-5}]$ at fixed $\beta$.
    Colored dashed lines represent the trends predicted by equation \eqref{eq:gammamax_largePe_largepsi} in (\textit{a}) and (\textit{c}), and by equation \eqref{eq:kmax_largePe_largepsi} in (\textit{b}) and (\textit{d}), using the best-fit coefficients listed in Table \ref{tab:coefficients_from_fit}.
    }
    \label{fig:max_values_vs_Gamma_and_beta}
\end{figure}

In Figure \ref{fig:max_values_vs_Gamma_and_beta} we plot $\gamma_{\max}$ (panels (\textit{a}) and (\textit{c})) and $k_{\max}$ (panels (\textit{b}) and (\textit{d})) as functions of $\Gamma$ and $\beta$ while setting $\Peclet = 10^3$. For all panels, the comparison with the prediction from equations \eqref{eq:gammamax_and_kmax_largePe_largepsi} is shown for all $\Gamma$ and for small enough $\beta$.

\begin{table}
  \begin{center}
  \def~{\hphantom{0}}
  \begin{tabular}{ll}
    Coefficients for $\gamma_{\max}$ & Coefficients for $k_{\max}$ \\[3pt]
    $a_\gamma = -0.767 \pm 0.002$ & $a_k = -0.896 \pm 0.001$ \\
    $b_\gamma = -3.63 \pm 0.02$  & $b_k = -1.701 \pm 0.006$ \\
  \end{tabular}
  \caption{Coefficients obtained from Eqs. \eqref{eq:gammamax_largePe_largepsi} and \eqref{eq:kmax_largePe_largepsi}, performing fit of the data shown in Figure \ref{fig:max_values_vs_Gamma_and_beta} for $\beta \in [10^{-5}, 10^{-2.5}]$. Uncertainties reflect one standard deviation, estimated from the covariance matrix of the fit.}
  \label{tab:coefficients_from_fit}
  \end{center}
\end{table}

Looking at panels (\textit{a}) and (\textit{b}), we note that both $\gamma_{\max}$ and $k_{\max}$ increases monotonically with $\Gamma$. For the growth rate, we note that the characteristic time of the cooling mechanism of the fluid through the plates is $\tau_c \propto1/\Gamma$, so a mechanism controlled by that process will have a characteristic rate scaling like $\gamma \sim 1/\tau_c \sim \Gamma$. Regarding $k_{\max}$, we expect that the typical perturbation wavelength scales with the characteristic length of the base state $1/\xi$ (also referred as thermal entry length). This scaling is consistent with previous findings \citep{Wylie1995, Morris1996}. Furthermore, from \eqref{eq:def_xi} we see that $\xi$ is a monotonically increasing function of $\Gamma$. Thus, the typical perturbation wavenumber grows with $\Gamma$.

On the other hand, in panel (\textit{c}) we observe that $\gamma_{\max}$ decreases monotonically with increasing $\beta$ for all choices of $\Gamma$, eventually becoming negative. This confirms that stronger viscosity contrasts promote the onset and growth of instabilities \citep{Holloway2005}. Moreover, from panel (\textit{d}) we see that $k_{\max}$ scales as well with the inverse of $\beta$, exhibiting a distinct linear trend on a semi-logarithmic scale, at least for $\beta\lesssim 10^{-2.5}$. Higher viscosity contrasts make the most unstable wavelengths progressively shorter.

For all panels in Figure \ref{fig:max_values_vs_Gamma_and_beta}, as $\Peclet \gg 1$, the comparison with the prediction from Eq.s \eqref{eq:gammamax_and_kmax_largePe_largepsi} is shown for all $\Gamma$ and for small enough $\beta$, namely in the interval $\beta \in [10^{-5}, 10^{-2.5}]$. A linear fit in semi-logarithmic scale returns the coefficients reported in Table \ref{tab:coefficients_from_fit}, showing a good agreement with the analytical prediction.

Moreover, for sufficiently large $\beta$, we observe that $\gamma_{\max}$ remains negative regardless of $\Peclet$ and $\Gamma$. We will discuss this aspect further in \S\,\ref{sec:betac}.

\begin{figure}
    \centering
    \includegraphics[width=1.\linewidth]{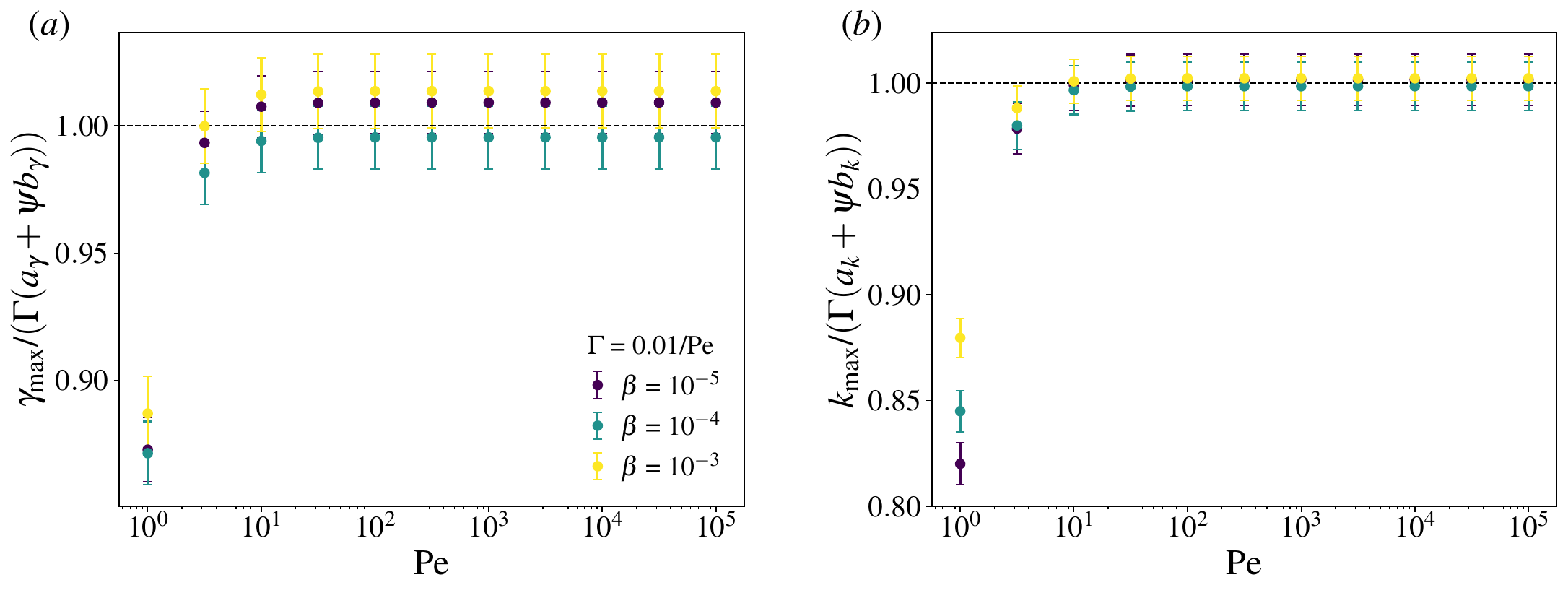}
    \caption{Rescaled maximum growth rate $\gamma_{\rm max}/(\Gamma(a_\gamma\psi+b_\gamma))$ (panel (\textit{a})) and corresponding wavenumber $k_{\rm max}/(\Gamma(a_k\psi+b_k))$ (panel (\textit{b})) as functions of the Péclet number $\Peclet \in [1, 10^{6}]$. For each dot we set $\Gamma = 0.01/\Peclet$, while different values of $\beta$ are indicated by different colors. The error bars represent standard deviations obtained via linear error propagation from uncertainties in $\gamma_{\rm max}, a_\gamma$ and $b_\gamma$ (and analogously for $k_{\rm max}, a_k$ and $b_k$). The dashed horizontal lines in panel (\textit{a}) and (\textit{b}) represent the high Péclet number asymptotic limit $\gamma_{\rm max}/(\Gamma(a_\gamma\psi+b_\gamma)) = 1$ and $k_{\rm max}/(\Gamma(a_k\psi+b_k)) = 1$, respectively.
    }
    \label{fig:max_values_vs_Pe_rescaled}
\end{figure}

To test the validity of the high-$\Peclet$ limit, we plot the rescaled quantities $\gamma'_{\rm max} = \gamma_{\rm max}/(\Gamma(a_\gamma\psi+b_\gamma))$ and $k'_{\rm max} = k_{\rm max}/(\Gamma(a_k\psi+b_k))$ for different values of $\Peclet$, as shown in Figure \ref{fig:max_values_vs_Pe_rescaled}. We can see that the rescaled growth rate and wavenumber are compatible, within the error, with the asymptotic limit $\gamma'_{\rm max}=1$ and $k'_{\rm max}=1$, respectively, as long as the Peclet number is sufficiently high. This behavior holds for different choices of $\Gamma$ and $\beta$, as long as $\Peclet\, \Gamma \ll 1$ and $\beta\ll 1$, with deviations visible for $\Peclet \lesssim 10$.

\subsection{Critical viscosity ratio}
\label{sec:betac}
An important distinction between thermally driven fingering and the classical Saffman–Taylor instability lies in the existence of a critical viscosity ratio $\beta_c < 1$ for the onset of thermal fingering. This threshold has been previously found in theoretical and numerical analyses \citep{Helfrich1995, Wylie1995, Diniega2013, Taylor-West2025} as well as in experimental observations \citep{Holloway2005}. In contrast, the Saffman–Taylor instability is triggered whenever the viscosity of the displacing fluid is lower than that of the displaced fluid, i.e. when the viscosity ratio is less than one \citep{SaffmanTaylor1958}. Similarly, in the case of miscible viscous fingering, instabilities arise as soon as the mobility ratio, governed by the concentration-dependent viscosity, falls below unity, even in the presence of nonlinear chemical reaction terms \citep{Tan1986, Tan1988, DeWit1999}.

As seen in Section \ref{sec:lsa_gammamax_kmax}, the growth rate $\gamma_{\max}$ saturates at a constant value for sufficiently high $\Peclet$, given $\Gamma$ and $\beta$. We therefore expect the critical viscosity ratio $\beta_c$, to converge to a specific value as $\Peclet \gg 1$. We determine the critical point by performing a quadratic fit of the data around the expected root. Specifically, we fit the pairs $(\log\beta, \gamma_{\max})$ shown in Fig. \ref{fig:max_values_vs_Gamma_and_beta}(c) over the interval $\beta \in [10^{-2.5}, 10^{-1.5}]$, using all datasets corresponding to the different values of $\Gamma$ in a single, simultaneous fit. This procedure yields the critical value $\psi_c = -\log\beta_c$, where
\begin{equation}
    \psi_c = 4.40 \pm 0.01.
    \label{eq:psi_c}
\end{equation}
Physically, this result implies that when advection dominates over thermal diffusion, the system becomes unstable only if the viscosity ratio satisfies $\beta < \beta_c$ where $\beta_c = (1.23 \pm 0.02)\cdot 10^{-2}$. This value can be compared with the one found numerically by \cite{Helfrich1995} when considering uniform flow injection, namely $\psi_c \simeq 4.27$.

\subsection{Dispersion relation for vanishing viscosity ratio}
\label{sec:disp_rel_high_psi}
In \S\,\ref{sec:lsa_gammamax_kmax}, we have focused on the behavior of the maxima, since they correspond to the most dangerous instabilities. However, it is possible to characterize the full dispersion relation. In the limit of large $\Peclet$, we find that the growth rate can be expressed in terms of a dispersion relation that is independent of both $\Gamma$ and $\beta$. 

Introducing the rescaled growth rate $\tilde{\tilde{\gamma}} = (\gamma/\Gamma + 1)/\psi$ and the corresponding rescaled wavenumber $\tilde{\tilde{k}} = k/(\Gamma\psi)$, we obtain, in the limit of large $\psi = -\log\beta$,
\begin{equation}
    \tilde{\tilde{\gamma}}(\tilde{\tilde{k}},\psi)
    = f_0(\tilde{\tilde{k}}) + \psi^{-1} f_1(\tilde{\tilde{k}}).
    \label{eq:tilde_tilde_gamma}
\end{equation}
Here, $f_0$ and $f_1$ are functions that do not depend on any global parameter. In particular, $f_0$ represents the rescaled dispersion relation in the limit $\psi \to \infty$ (i.e., $\beta \to 0$), while $f_1$ provides a correction that is linear in $1/\psi$. The quantity rescaling and the high-$\psi$ perturbation shown here are suggested by the parameter regrouping discussed in the Appendix \ref{sec:app:largePe_smallbeta}.

\begin{figure}
    \centering
    \includegraphics[width=1.05\linewidth]{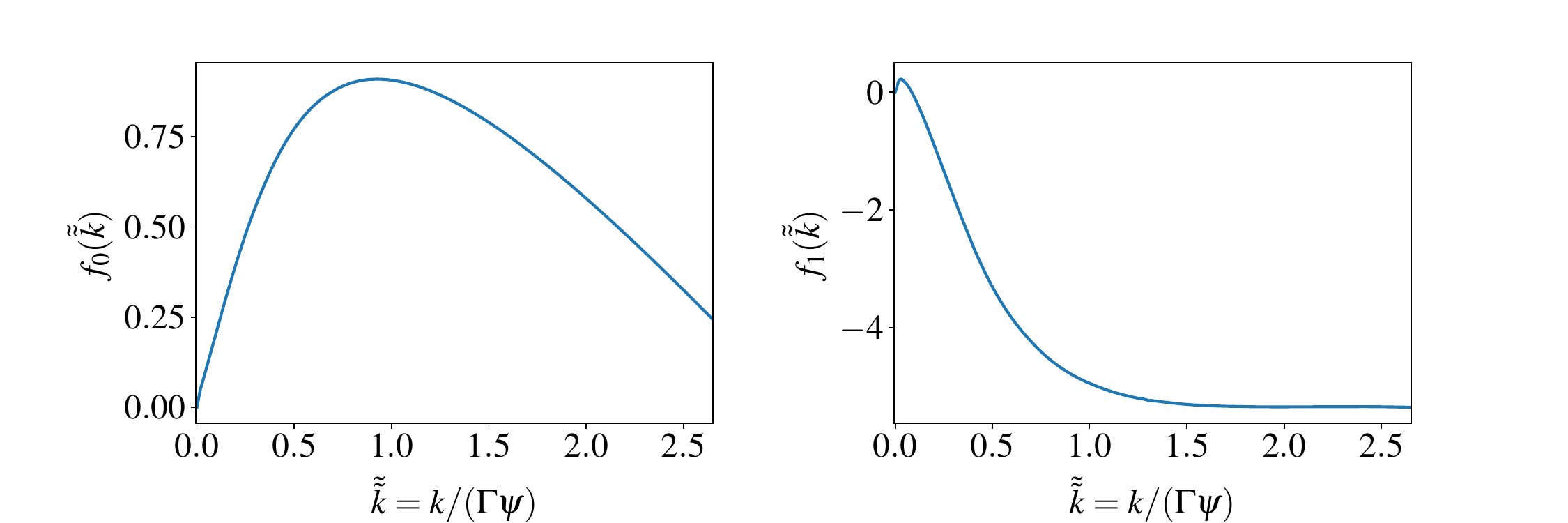}
    \caption{Numerical reconstruction of $f_0$ and $f_1$ as functions of the rescaled wavenumber $\tilde {\tilde k} = k/(\Gamma\psi)$ by using Eqs. \eqref{eq:f_0_f_1}.}
\label{fig:f_0_f_1}
\end{figure}

Both $f_0$ and $f_1$ can be extracted numerically by computing the rescaled dispersion relation $\tilde{\tilde{\gamma}}(\tilde{\tilde{k}},\psi)$ for two different values of $\psi$, called $\psi_l$ and $\psi_r$. Using \eqref{eq:tilde_tilde_gamma}, we obtain
\begin{align}
    f_1(\tilde{\tilde{k}}) &= \frac{\tilde{\tilde{\gamma}}(\tilde{\tilde{k}};\psi_l)
    - \tilde{\tilde{\gamma}}(\tilde{\tilde{k}};\psi_r)}
    {1/\psi_l - 1/\psi_r}, \\
    f_0(\tilde{\tilde{k}}) &= \tilde{\tilde{\gamma}}(\tilde{\tilde{k}};\psi_l)
    - \psi_l^{-1} f_1(\tilde{\tilde{k}}),
    \label{eq:f_0_f_1}
\end{align}
In Figure \ref{fig:f_0_f_1}, we show the numerical reconstruction of both $f_0$ and $f_1$ obtained by generating two dispersion relations with $\beta_l = e^{-\psi_l} = 10^{-10.25}$ and $\beta_r = e^{-\psi_r} = 10^{-10}$ ($\psi_l \approx 23.6$ and $\psi_r \approx 23$, respectively), while keeping $\Gamma = 10^{-5}$ and $\Peclet = 10^3$; for both curves, the $(\gamma, k)$ couples were generated in the range $k\in [0,\,6.25\cdot10^{-4}]$ at intervals of $2\cdot10^{-6}$.

From the resulting $f_0$ and $f_1$ it comes that, within numerical error, $f_0(0) = f_1(0) = 0$, as visible in Fig. \ref{fig:f_0_f_1}. This implies that $\gamma(k=0) = -\Gamma$ for all combinations of global parameters, provided that $\Peclet\, \Gamma \ll 1$ and both $\Peclet$ and $\psi$ are sufficiently large. A perturbation of vanishing wavenumber keeps the flow uniform, so the $y$-independent exponential decay of the temperature along the main flow direction is preserved. For a set position $x$, the temperature decays towards the base state as $T(t)=T_0(x)(1+\epsilon e^{-\Gamma t})$, being $\epsilon$ the perturbation intensity, showing that $\Gamma$ constitutes the exponential decay rate of the perturbation. We can also check this by seeing that the decay length of the exponential tail of the perturbation $1/\Lambda$, where $\Lambda$ is defined in Eq. \eqref{eq:Lambda}, diverges when setting $k = 0$ and $\gamma = - \Gamma$.

Another consequence of Eq. \eqref{eq:tilde_tilde_gamma} is that we can write both the rescaled maximum growth rate $\tilde {\tilde \gamma}_{\max}$ and wavenumber $\tilde {\tilde k}_{\max}$ as a linear combination of two independent quantities respectively, i.e.
\begin{align}
    &\tilde {\tilde k}_{\max} = \tilde{\tilde k}_{\max,0} + \psi^{-1}\tilde{\tilde k}_{\max,1},\\
    &\tilde {\tilde \gamma}_{\max} = \tilde {\tilde \gamma}_{\max,0} + \psi^{-1} \tilde {\tilde \gamma}_{\max,1}.
\end{align}
$\tilde{\tilde k}_{\max,0}$ is the value of $\tilde{\tilde k}$ that realizes the maximum of $f_0$, while $\tilde{\tilde k}_{\max,1}$, $\tilde {\tilde \gamma}_{\max,0}$ and $\tilde {\tilde \gamma}_{\max,1}$ can be obtained keeping the terms linear in $\psi^{-1}$ in both $\tilde {\tilde \gamma}(\tilde {\tilde k}_{\max})$ and $\tilde {\tilde \gamma}'(\tilde {\tilde k}_{\max})$, and furthermore imposing $\tilde {\tilde \gamma}'(\tilde {\tilde k}_{\max}) = 0$. It comes that $\tilde{\tilde k}_{\max,1} = -f'_1(\tilde{\tilde k}_{\max,0})/f''_0(\tilde{\tilde k}_{\max,0})$, $\tilde {\tilde \gamma}_{\max,0}= f_0(\tilde {\tilde k}_{\max,0})$ and $\tilde {\tilde \gamma}_{\max,1}= f_1(\tilde {\tilde k}_{\max,0})$. Back to the non-rescaled quantities,
\begin{align}
    & k_{\max} = \Gamma (\psi\tilde{\tilde k}_{\max,0} + \tilde{\tilde k}_{\max,1}) = \Gamma(a_k \log \beta + b_k), \label{eq:k_max_from_rescaling} \\
    & \gamma_{\max} = \Gamma (\psi f_0(\tilde {\tilde k}_{\max,0}) + f_1(\tilde {\tilde k}_{\max,0}) - 1) = \Gamma(a_\gamma \log \beta + b_\gamma), \label{eq:gamma_max_from_rescaling}
\end{align}
retrieving a linear relationship of both $k_{\max}$ and $\gamma_{\max}$ with $\Gamma$ and $\beta$, identical to the one discussed in \S\ref{sec:lsa_gammamax_kmax}. The numerical coefficients correspond to $a_k = -\tilde{\tilde k}_{\max,0}$, $b_k = \tilde{\tilde k}_{\max,1}$, $a_\gamma = -f_0(\tilde{\tilde k}_{\max,0})$ and $b_\gamma = f_1(\tilde{\tilde k}_{\max,0}) - 1$. We locate $(\tilde{\tilde k}_{\max,0},\,f_0(\tilde{\tilde k}_{\max,0}))$ performing a quadratic fit around the expected maximum of $f_0$, and we estimate the value of $f_1$ and its first derivative at $\tilde{\tilde k} = \tilde{\tilde k}_{\max,0}$ doing a quadratic fit of $f_1$ in an interval around $\tilde{\tilde k}_{\max,0}$. The values obtained are reported in Table \ref{tab:coefficients_from_rescaling}.

Comparing them with the same coefficients reported in Table \ref{tab:coefficients_from_fit}, we can see that $a_k$ and $b_k$ show a good compatibility within the uncertainty reported. Furthermore, the two methods are in moderate agreement for $a_\gamma$ (relative difference of $\approx 15 \%$), while $b_\gamma$ differs more substantially ($\approx 60\%$). We attribute this discrepancy to the fact that the range of $\beta$ values for the global-fit presented in Section \S\ref{sec:lsa_gammamax_kmax}, i.e. $\beta \in [10^{-5},10^{-3}]$, does not fully reach the asymptotic regime probed in the method shown in this Section, making especially the intercept $b_\gamma$ sensitive to sub-leading corrections. Nevertheless, the global fit shows that the linear relationship between both $(\log\beta, k_{\max})$ and $(\log\beta, \gamma_{\max})$ already holds remarkably well even in the moderate-$\psi$ regime considered. The asymptotic trend sets in early, even though the fitted coefficients do not yet match their asymptotic values.

\begin{table}
  \begin{center}
  \def~{\hphantom{0}}
  \begin{tabular}{ll}
    Coefficients for $\gamma_{\max}$ & Coefficients for $k_{\max}$ \\[3pt]
    $a_\gamma = -0.93 \pm 0.01$ & $a_k = -0.91 \pm 0.01$ \\
    $b_\gamma = -5.83 \pm 0.04$  & $b_k = -1.66 \pm 0.06$ \\
  \end{tabular}
  \caption{Coefficients obtained from Eqs. \eqref{eq:k_max_from_rescaling}–\eqref{eq:gamma_max_from_rescaling} sampling the functions $f_0$ and $f_1$ shown in Fig. \ref{fig:f_0_f_1}. Uncertainties are derived via linear propagation from a local quadratic fit of both $f_0$ and $f_1$ around $\tilde{\tilde k}_{\max,0}$.}
  \label{tab:coefficients_from_rescaling}
  \end{center}
\end{table}

\section{Comparison between linear stability analysis and nonlinear simulations}
\label{sec:compare}
\begin{figure}
    \centering
    \includegraphics[width=1.\linewidth]{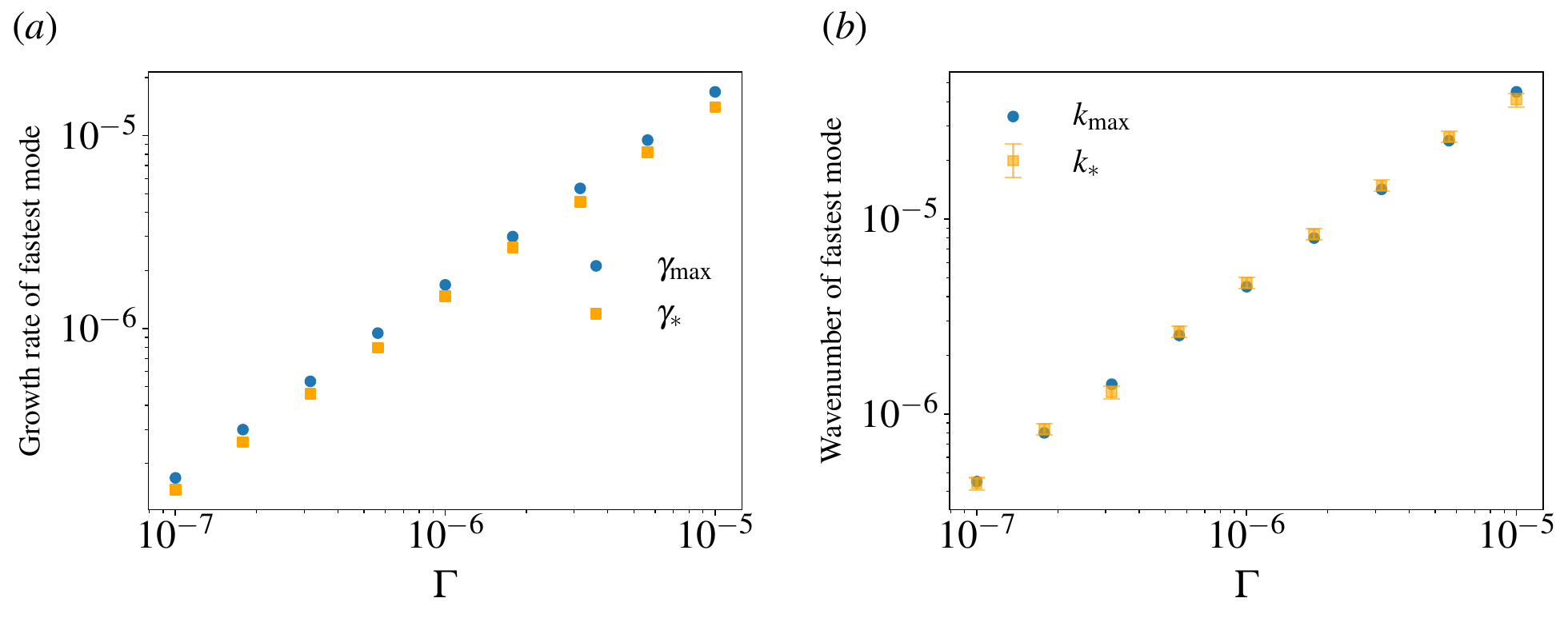}
    \caption{(\textit{a}) Growth rate of the fastest mode and (\textit{b}) corresponding wavenumber as functions of the Péclet number $\Gamma$ in the range $[10^{-7}, 10^{-5}]$, for $\Peclet = 10^3$ and $\beta = 10^{-3}$. Results from linear stability analysis are shown as blue circles, while values extracted from nonlinear simulations are shown as orange squares.
    The error bars on $k_*$ reflect the finite system size in full simulations and are equal to $\pm 2\pi/L_y$, where $L_y$ is the transverse domain length.}
\label{fig:max_values_vs_Pe_compare}
\end{figure}

To check the predictive power of the linear theory, we compare its results with those obtained from nonlinear simulations. Figure \ref{fig:max_values_vs_Pe_compare} shows the growth rate of the fastest mode and its wavenumber as functions of the cooling rate $\Gamma$, for fixed values $\Peclet = 10^3$ and $\beta = 10^{-3}$.

The linear stability predictions $\gamma_{\max}$ and $k_{\max}$ are compared with the corresponding values $\gamma_*$ and $k_*$ extracted from the full simulations, as described in \S\,\ref{sec:full_random}. The two sets of results are in good agreement across the entire range of $\Gamma$ considered, i.e.\ $\gamma_{\max} \approx \gamma_*$ and $k_{\max} \approx k_*$.
The error bars of the measured wavenumber $k_* = 2\pi n_f/L_y$ shown in Fig. \ref{fig:max_values_vs_Pe_compare}(\textit{b}) are estimated by taking into account the finite size $L_y$ of the domain in the full simulations. In particular, we assume that the error $\Delta k_* = 2\pi \Delta n_f/L_y = \pm 2\pi/L_y$ reflects that the error in the number of fingers $n_f$ may be up to $|\Delta n_f| = 1$, due to $L_y$ not being a multiple of the theoretical optimal wavelength $2\pi/k_{\rm max}$. We can see that most unstable wavenumber computed from the linearized model is within these error bars.

We also note that the measured growth rate, shown in Fig. \ref{fig:max_values_vs_Pe_compare}(\textit{a}) is systematically slightly below the maximal growth rate based on the linearized model. This behavior can be explained by finite-size effects; as the full model is based on a system size $L_y$, the actual wavenumber measured, $k_*$, will be somewhat away from the theoretical optimum $k_{\rm max}$ (consistent with Fig. \ref{fig:max_values_vs_Pe_compare}(\textit{a})) and thus the actual growth rate measured, $\gamma_*$, will be somewhat below the theoretical maximum $\gamma_{\rm max}$. This emphasizes that when comparing to experiments or simulations in finite systems, $\gamma_{\rm max}$ should be viewed as an upper bound for the growth rate.

The agreement between the full and linearized models confirms that the linear stability analysis accurately captures the dominant features of the instability during its early evolution, including both the characteristic growth rate and the selected wavelength.

\section{Discussion and conclusion}
\label{sec:discussion}

In this work, we have investigated a thermo–viscous instability that arises when a hot, low–viscosity fluid is injected into a narrow channel conducting heat to a colder environment. Larger viscosity contrasts promote faster growth and shorter wavelengths, as in classical fingering phenomena, but the physical mechanism differs fundamentally from both immiscible and miscible Saffman–Taylor–type instabilities. Here the stabilizing effect is controlled by heat transfer: cooling of the injected fluid reduces the viscosity contrast and competes with amplification induced by advection.

A key novelty of our analysis is the focus on the long–time, small Biot number regime, in which heat can diffuse through the solid plates and the fluid layer such that the cross–gap temperature becomes nearly uniform. In this asymptotic limit, cross–gap averaging yields a depth-averaged model that includes a Taylor–dispersion contribution and captures the fully coupled feedback between velocity and temperature.

Through linear stability analysis, we show that the competition between the fluid advection, the in-plane thermal diffusion, the hydrodynamic (Taylor) dispersion along the flow direction and the out-of-plane heat loss through the boundary walls, gives rise to a dispersion relation which relies on the three global parameters: Péclet number, cooling rate, and viscosity ratio.

We demonstrate that the maximum growth rate and most unstable wavenumber of the instability scale with a characteristic time $\Gamma^{-1}$ and a characteristic length scale $\xi^{-1}$, respectively. The former characterizes out-of-plane heat flux through the walls, whereas the latter results from a combination of in-plane molecular diffusion, Taylor dispersion, and heat flux through the walls. Remarkably, these scales are physically related to the solution of two homogeneous problems respectively. The first problem corresponds to that of a stationary homogeneous reservoir under hydrostatic conditions, for which equation \eqref{eq:advdiff} can be simplified to $\mathrm{d}T /\mathrm{d}t = - \Gamma T$, which, with the initial condition $T(t=0)=1$, leads to the solution $T(t)=e^{-\Gamma t}$, where $\Gamma$ gives the exponential decay rate of the temperature. The second is the steady state one-dimensional problem shown in \S\,\ref{sec:model_base_state}, whose base-state solution is $T_0(x) = e^{-\xi x}$, shown in Eq. \eqref{eq:base_state}, where $1/\xi$ is the temperature decay length.

We furthermore find that, in the asymptotic regime of large $\Peclet$ and small $\beta$, both the maximum growth rate and the the corresponding wavelength are governed by the heat transfer coefficient and the viscosity ratio alone. We combine equations \eqref{eq:gammamax_and_kmax_largePe} and \eqref{eq:gammamax_and_kmax_largePe_largepsi} in a dimensional form, writing the maximal growth rate $\gamma_{\rm max}^{\rm dim}$ and the most unstable wavelength $k_{\rm max}^{\rm dim}$ in terms of the physical quantities characterizing the problem:
\begin{subequations}
\begin{align}
    \gamma_{\max}^{\rm dim} &\overset{U h\gg \kappa, \ \beta\ll 1}{\simeq}  \frac{H_{\rm ov}}{\rho c h} (a_\gamma \log\beta + b_\gamma), \label{eq:gammamax_largePe_dim} \\
    k_{\max}^{\rm dim}  &\overset{U h\gg \kappa, \ \beta\ll 1}{\simeq} \frac{H_{\rm ov}}{\rho c h U } (a_k\log\beta + b_k). \label{eq:kmax_largePe_dim}
\end{align}\label{eq:gammamax_and_kmax_largePe_dim}\end{subequations}
We recall that $U$ is the mean flow velocity, $h$ is the half-height of the Hele-Shaw cell, $\rho$ is its density, $c$ is its specific heat capacity, $H_{\rm ov}$ the overall heat transfer coefficient. Estimations for the numerical constants $a_i, b_i$, with $i = {\gamma,k}$, were realized in two different ways: (i) performing a linear fitting of the measured values of $\gamma_{\max}$, $k_{\max}$ versus $\Gamma$ and $\log\beta$ (ref. \S\,\ref{sec:lsa_gammamax_kmax}, values shown in Table \ref{tab:coefficients_from_fit}); (ii) sampling the maximum of the dispersion relation numerically reconstructed in the limit of high $\psi$ (ref. \S\,\ref{sec:disp_rel_high_psi}, values shown in Table \ref{tab:coefficients_from_rescaling}). While method (ii) more directly probes the $\psi \to \infty$ limit, giving a more accurate estimation of the theoretical asymptotic values, method (i) gives effective coefficients appropriate for the moderate-$\psi$ regime actually accessible in experiments (see \S\,\ref{sec:discussion_applications}).

We show that the stability of the system is governed by a critical $\beta_c$, such that the system is unconditionally stable if $\beta>\beta_c$, in agreement with previous findings. For large Péclet numbers, we found $\psi_c = -\log\beta_c \approx 4.40$.

We note that, in the limit $\Peclet \Gamma \ll 1$ and $\Peclet \gg 1$, our model reduces to the one studied by \cite{Helfrich1995}. However, their analysis relies on different boundary conditions at the wall: they impose a constant temperature, appropriate for times shorter than the characteristic cross-gap diffusion time. In addition, their averaging procedure assumes a strongly non-uniform temperature profile together with a uniform mobility across the gap. This internal inconsistency nevertheless results in a model formally identical to ours, though its physical interpretation is remarkably different. Finally, we advance the analysis by deriving new results for the dispersion relations and their maximal values.

Further extensions of the present model could take into account viscous friction heating \citep{Costa2005}, which, in the context of viscous fingering, may contribute in stabilizing the system due to the increase in temperature and consequent reduction in mobility in the displaced fluid, decreasing the viscosity contrast between the invading and displaced fluid \citep{Norouzi2019}. Also, taking into account yield stress in the model might be of interest, as, for example, magmatic flow can crystallize and transition to a Bingham flow \citep{Dragoni2002, Castruccio2013}. In this case, the instability should be favored, since the cooled fluid would be slowed even more than predicted by the Newtonian flow model.

\subsection{Application to geological and laboratory settings}
\label{sec:discussion_applications} 
Our results indicate that the thermo-viscous channelization mechanism identified here should be observable both in laboratory experiments and in natural magmatic systems. In the existing laboratory studies \citep{Whitehead1991, Wylie1999, Holloway2005, Nagatsu2009} the wall temperature is kept constant and therefore the system is at $\Biot \sim 1$. In contrast, our mechanism requires a small Biot number, which can be realized in a Hele–Shaw type cell consisting of two conducting plates (e.g.\ stainless steel or aluminum, $k_s\sim 15$--$200\,\mathrm{W\,m^{-1}K^{-1}}$) separated by a uniform gap $h=0.2$--$0.5\,\mathrm{mm}$. In such a configuration, the external reservoir temperature is controlled, but the solid-fluid interface temperature evolves consistently through heat exchange with the surroundings. Using a glycerol-water mixture or a silicone oil sensitive to temperature, a temperature contrast of $60$--$100^{\circ}\mathrm{C}$ produces viscosity ratios $\beta=\mu_h/\mu_c\sim 10^{-3}$--$10^{-2}$ \citep{Cheng2008, Venczel2021}, corresponding to $\psi = -\log\beta \approx 4.5$--$7$, safely above the instability threshold \eqref{eq:psi_c}. For typical fluid properties ($\rho\approx 10^3\,\mathrm{kg\,m^{-3}}$, $c\approx 3$--$4\times10^3\,\mathrm{J\,kg^{-1}K^{-1}}$) and imposed inlet velocities $U\sim 10^{-4}$--$10^{-3}\,\mathrm{m\,s^{-1}}$, the thermal coupling coefficient becomes $\Gamma = H_{\rm ov}/(\rho c U)\sim 10^{-2}$--$10^{-1}$ for realistic plate-fluid heat transfer coefficients $H_{\rm ov}\sim 20$--$200\,\mathrm{W\,m^{-2}K^{-1}}$. This yields $\Biot = H h/k_s \lesssim 0.05$, within the small $\Biot$ regime. The predicted characteristic time and wavelength, $t_{\mathrm{char}}=1/\gamma^{\mathrm{dim}}_{\max}$ and $\lambda_{\mathrm{char}}=2\pi/k^{\mathrm{dim}}_{\max}$, with $\gamma^{\mathrm{dim}}_{\max}$ and $k^{\mathrm{dim}}_{\max}$ computed from Eqs. \eqref{eq:gammamax_and_kmax_largePe_dim}, are in the range of minutes to hours and $5$--$20\,\mathrm{mm}$, respectively. This makes the instability directly observable in a $10$--$20\,\mathrm{cm}$ cell. Although the Péclet number would be moderate ($\Peclet\sim 1$--$5$), our simulations show that the thermo–viscous instability persists in this regime.

The same physical conditions can arise in magmatic systems. If basaltic magma is confined in millimeter–scale fissures, which are well documented despite more common meter–scale dykes \citep{Krumbholz2014}, its material properties are favorable for thermo–viscous channelization predicted by our theory. Basaltic magma exhibits density $\rho\approx 2700\,\mathrm{kg\,m^{-3}}$, heat capacity $c\approx 10^3\,\mathrm{J\,kg^{-1}K^{-1}}$, thermal conductivity $k_f\approx 0.5$--$1\,\mathrm{W\,m^{-1}K^{-1}}$, and viscosity contrasts of $10^2$--$10^4$ over typical  magmatic temperature ranges (so $\psi\approx 4$--$9$) \citep{Diniega2013}. Host rocks have thermal properties $k_s\approx 1$--$3~\mathrm{W\,m^{-1}K^{-1}}$ and $\kappa_s\sim 10^{-6}\,\mathrm{m^2\,s^{-1}}$. For fissure half widths $h\sim 1$--$5~\mathrm{mm}$, and assuming magma velocities $U\sim 10^{-3}$--$10^{-2}\,\mathrm{m\,s^{-1}}$, the resulting thermal coupling parameter is $\Gamma\sim 10^{-5}$--$10^{-4}$ for realistic coefficients of wall cooling $H_{\rm ov}\sim 1$--$10\,\mathrm{W\,m^{-2}K^{-1}}$. Substituting these values into our asymptotic predictions returns characteristic growth times of hours to days and wavelengths from tens of meters to kilometers, consistent with the scales of observable in–plane flow localization and structural segmentation in dykes. These results indicate that thermo–viscous channelization at small $\Biot$ may provide a plausible mechanism for the emergence of preferential flow paths in narrow magmatic conduits.

A related regime occurs in thermo–viscous lubrication films in mechanical systems, where silicone or polyalphaolefin (PAO) oils confined in micrometer-scale gaps experience strong temperature–dependent viscosity reduction under shear heating. Such films typically operate at small Biot numbers and moderate to large Peclet numbers, and can develop localized thermal channels and streaks \citep{Schmid2013, Khonsari2017}. Although in general geometrically different, these systems share the same coupling mechanism between heat transfer, advection, and viscosity, possibly expanding the applicability of the instability studied here. In our simulations, the constant inlet flow rate boundary condition provides a physically realistic representation of the conditions of oil supply typically present in such lubrication systems.

\begin{acknowledgments}
\textbf{Funding.}
This work was supported by the Research Council of Norway through its Center of Excellence funding scheme (PoreLab, project 262644), and through its Researcher Project for Young Talents programme (M4, project 325819, and UNLOC, project 345008).

\textbf{Acknowledgments.}
We acknowledge the contribution of Beatrice Baldelli who performed early exploratory simulations of our problem. We are grateful to Hans Jørgen Kjøll, Olivier Galland, and Renaud Toussaint for fruitful discussions.

\textbf{Data availability statement.}
The code and scripts used to generate the results presented here will be made available on GitHub upon publication \cite{CoolingGit2025}.

\textbf{Author contributions.}
\textbf{F. Lanza:} Data curation (lead); Formal analysis (equal); Investigation (lead); Software (equal); Validation (lead); Visualization (lead); Writing - original draft (lead).
\textbf{G. Linga:} Conceptualisation (equal); Formal analysis (equal); Methodology (equal); Software (equal); Validation (supporting); Writing - original draft (supporting); Writing - review \& editing (equal);
\textbf{F. Barras:} Conceptualisation (equal); Formal analysis (equal); Methodology (supporting); Writing - review \& editing (equal);
\textbf{E. G. Flekkøy:} Conceptualisation (equal); Formal analysis (supporting); Funding acquisition (equal); Methodology (equal); Project administration (lead); Supervision (lead); Writing - review \& editing (supporting).
\end{acknowledgments}

\appendix

\section{Cross-gap averaging}
\label{sec:averaging}

We recall the governing 3D equations for an incompressible Newtonian fluid with temperature-dependent viscosity $\mu(T)$
\begin{align}
    \div \vec u &= 0, \label{eq:supp_mass_cons} \\
    \div [ 2 \mu (T) \gv \varepsilon ] &= \grad p,  \label{eq:supp_momentum_cons} \\
    \pd T t + \vec u \bdot \grad T - \kappa \laplacian{T} &= 0, \label{eq:supp_advdiff}
\end{align}
being $\vec u = (u_x,u_y,u_z)$. In a slit of half gap $h$ and in-plane length scale $L$ with $h/L\ll1$, impermeable walls, and low-Reynolds-number flow, lubrication theory gives $|(\partial_x,\partial_y)(u_x,u_y)|\sim |(u_x,u_y)|/L \ll |(u_x,u_y)|/h$ and $u_z=O(Uh/L)$ from continuity. We also assume that in-plane variations of viscosity are slow, so $|\nabla_\parallel \mu| \sim \mu/L \ll \mu/h$
and terms involving in-plane viscosity gradients are $O(h/L)$ smaller than the leading transverse diffusion. At leading order, the momentum balance reduces to
\begin{equation}
\pdx{z}\!\left(\mu(T)\,\pdx{z} (u_x,u_y)\right) = \nabla_\parallel p, \qquad \pd{p}{z} = 0. \label{eq:supp_momentum_cons_lub}
\end{equation}
with $\nabla_\parallel=(\partial_x,\partial_y)$.
If $T$ (hence $\mu$) is uniform in $z$, integration of \eqref{eq:supp_momentum_cons_lub} with no slip at $z=\pm h$ yields the parabolic in-plane profile
\begin{equation}
  \boldsymbol u_\parallel(x,y,z)= -\,\frac{1}{2\mu}\,\nabla_\parallel p\,\big(h^2-z^2\big)
  = \frac{3}{2}\,\overline{\boldsymbol u}_\parallel\!\left(1-\frac{z^2}{h^2}\right),
  \qquad 
  \overline{\boldsymbol u}_\parallel = -\,\frac{h^2}{3\mu}\,\nabla_\parallel p,
  \label{eq:supp_parabolic}
\end{equation}
where $\overline{(\cdot)}$ denotes the cross-gap average.

To leading order $\partial_z p=0$, so $p=p_\parallel(x,y)$ is $z$-uniform. Averaging \eqref{eq:supp_momentum_cons_lub} across $z\in[-h,h]$ and using \eqref{eq:supp_parabolic} gives the depth-averaged Stokes law
\begin{equation}
  -\,\frac{3}{h^2}\,\mu(T_\parallel)\, \overline{\vec u}_\parallel \;=\; \nabla_\parallel p_\parallel,
  \label{eq:avg-stokes}
\end{equation}
where $T_\parallel(x,y,t)=\overline{T}$ and we used that, for large times when $T$ is nearly uniform across the gap, $\mu(T)$ may be evaluated at $T_\parallel$ in the mobility. Averaging \eqref{eq:supp_mass_cons} with impermeable, stationary walls gives
\begin{equation}
  \nabla_\parallel\!\cdot\,\overline{\boldsymbol u}_\parallel \;=\; 0,
  \label{eq:avg-cont}
\end{equation}
For averaging \eqref{eq:supp_advdiff}, we will follow a derivation based on large-time asymptotics \citep{Liñán2020}, applied to our planar geometry. We introduce a moving frame aligned with the local mean flow, so that $\overline{\boldsymbol u}_\parallel=(\overline u,0)$ and we choose the following non-dimensional scales
\begin{equation}
    t' = \frac{t}{h^2/\kappa}, \quad x' = \frac{x-\overline u\, t}{h}, \quad z' = \frac{z}{h}, \quad p' = \frac{h}{\mu_c \overline u},\quad u' = \frac{u}{\overline u}, \quad \mu' = \frac{\mu}{\mu_c} \quad \Pecletlocal = \frac{\overline u h}{\kappa}
    \label{eq:supp_resc}
\end{equation}
In this moving frame, the temperature diffuses vertically. For $t' \gg 1$ (namely, times much greater than the out-of-plane diffusion time $h^2/\kappa$), thermal diffusion will cause the temperature to become uniform along $z$ at leading order, although temperature gradients along the $x$-direction still persist at sub-leading orders. If $t' \gg 1$, the temperature field will have spread along $x$ over a distance $x' \sim \sqrt{t'}$. To resolve the slow longitudinal spreading (Taylor dispersion), the further following rescaling can be introduced:
\begin{equation}
    \tau = \epsilon t', \quad  \zeta = \sqrt{\epsilon} x', \quad \epsilon \ll 1.
    \label{eq:supp_resc_eps}
\end{equation}
The advection-diffusion equation becomes
\begin{align}
    \epsilon \pd{T}{\tau} + \sqrt{\epsilon}\Pecletlocal\,(u'-1)\pd{T}{\zeta} - \epsilon \pdd{T}{\zeta} - \pdd{T}{z'} &= 0
    \label{eq:corr_advdiff}
\end{align}
We perform an expansion for the temperature and velocity fields in an asymptotic series, writing $T=T_0+\sqrt{\epsilon}\,T_1+\epsilon\,T_2+\cdots$ and $u'=u_0+\sqrt{\epsilon}\,u_1+\cdots$ with the base Poiseuille profile
\begin{equation}
  u_0(z')=\frac{3}{2}\big(1 - z'^2\big).
\end{equation}
By construction, the cross-gap average of $u'$ is $\overline u' = 1$. Being $\overline u_0 = 1$, it must be that $\overline u_{i}=0$ for $i\geq1$.\\
At leading order $O(\epsilon^0)$ we simply have
\begin{equation}
  \pdd{T_0}{z'} = 0,
\end{equation}
which, together with the symmetry condition $\pd{T_0}{z'}|_{z'=0} = 0$, gives $T_0 = T_0(\xi,\tau)$, namely $T_0$ is uniform in $z$.

At $O(\sqrt{\epsilon})$,
\begin{equation}
    \frac{\Peclet}{2} (1 - 3 z'^2)\pd{T_0}{\zeta} - \pdd{T_1}{z'} = 0
    \label{eq:supp_advdiff_resc2_ordersqrtepsilon}
\end{equation}
whose solution with the symmetry condition $\pd{T}{z'}|_{z'=0}=0$ gives
\begin{equation}
    T_1(\xi, z', \tau) =  \Pecletlocal \frac{z'^2}{4} \left(1 - \frac{z'^2}{2}\right)\pd{T_0}{\zeta} + T_{1}(z'\!=\!0)
    \label{eq:supp_T1_sol}
\end{equation}

Proceeding to $O(\epsilon)$, we have
\begin{equation}
    \pd{T_0}{\tau} + \Pecletlocal \, u_1\pd{T_0}{\zeta} - \pdd{T_0}{\zeta} + \frac{\Pecletlocal}{2} (1 - 3 z'^2)\pd{T_1}{\zeta} - \pdd{T_2}{z'} = 0.
    \label{eq:supp_advdiff_resc2_orderepsilon}
\end{equation}
Because 
\begin{equation}
\pd{T_0}{z'}|_{z'=1} = 0,\qquad \pd{T_1}{z'}|_{z'=1} = 0,
\end{equation}
the latter coming from deriving \eqref{eq:supp_T1_sol}, the Robin boundary condition at the wall \eqref{eq:bc_walls_3d} up at order $\epsilon$ reads
\begin{equation}
    \pd{T}{z'}\Bigg|_{z'=1} = \epsilon\pd{T_2}{z'}\Bigg|_{z'=1} = \frac{H_\mathrm{ov}h}{k_{\mathrm f}}\,(T_c - T_0).
    \label{eq:supp_advdiff_robin}
\end{equation}
Averaging Eq. \eqref{eq:supp_advdiff_resc2_orderepsilon} over the cell thickness, reminding that $\overline u_1 = 0$, gives 
\begin{equation}
    \pd{T_\parallel}{\tau} - \pdd{T_\parallel}{\zeta}\left(1 + \frac{2 \Pecletlocal^2}{105}\right) + \frac{\Biot}{\epsilon}(T_\parallel - T_c)  = 0
    \label{eq:supp_advdiff_resc2_orderepsilon_avg}
\end{equation}
where we recall the definition of the Biot number $Bi = H_\mathrm{ov} h/k_{\mathrm f}$. Note that consistency requires $Bi \sim \epsilon \ll 1$.
Returning to dimensional variables and any in-plane direction,
\begin{equation}
    \pd{T_\parallel}{t} + \vec u_\parallel \bdot \grad_\parallel T_\parallel - \grad_\parallel \bdot (\mathds{D} \grad_\parallel T_\parallel) + \frac{H}{h\rho c}(T_\parallel - T_c) = 0,
    \label{eq:supp_advdiff_2d}
\end{equation}
with the anisotropic Taylor-Aris dispersion tensor
\begin{equation}
   \mathds{D} = \kappa \left( \mathds{I} + \frac{2}{105}\left(\frac{h}{\kappa}\right)^2\vec u_\parallel \otimes \vec u_\parallel\right).
    \label{eq:supp_keff}
\end{equation}

\section{Numerical method}

\subsection{Finite element scheme for the fully coupled model}
\label{sec:num_meth_full}

We solve the fully coupled model \eqref{eq:full_model_2d} with a semi-implicit first–order operator–splitting scheme on a quadrilateral mesh. The computational domain is
\(\Omega=[0,L_x]\times[0,L_y]\). We use the scalar \(Q_1\) space \(S\) for both \(p\) and \(T\).
Periodic boundary conditions are imposed in the transverse (\(y\)) direction via
multi–point constraints (MPC), pairing \(y=0\) with \(y=L_y\) while excluding the corners to avoid double pinning.

Let the (clamped) mobility be
\begin{equation*}
m(T)=\mathrm{clip}\!\left(\exp(-(\ln\beta)\,T);\,m_{\min},\,m_{\max}\right), \qquad m_{\min}=10^{-8},\; m_{\max}=10^{8},
\end{equation*}
where the clamping function
\begin{equation*}
    \mathrm{clip}\!\left(x;\,x_{\min},\,x_{\max}\right) = 
    \begin{cases}
    x_{\min} & \text{for } x < x_{\mathrm{min}},\\[2pt]
    x & \text{for }x_{\min} \leq x \leq x_{\max},\\[2pt]
    x_{\max} & \text{for }x > x_{\max},
    \end{cases}
\end{equation*}
leaves values inside the admissible range unchanged, while truncating values that fall outside.\\
At time level \(t^n=n\,\Delta t\), the scheme reads:

\begin{enumerate}[label=(\roman*)]
\item \textbf{Initialization.} Set \(T^0(x,y)=\exp(-\xi x)\) from the base state \eqref{eq:base_state}.

\item \textbf{Pressure (Darcy) solve.} Find \(p^n\in S\) such that for all \(q\in S\),
\begin{equation}
\int_\Omega m(T^{n-1})\,\nabla p^n\cdot\nabla q\,\mathrm{d}x
=\int_{\Gamma_{\mathrm{in}}} U_{\mathrm{in}}(y,t^n)\,q\,\mathrm{d}s,
\quad p^n\big|_{\Gamma_{\mathrm{out}}}=0,
\label{eq:fem_p_smallGamma}
\end{equation}
with \(\Gamma_{\mathrm{in}}=\{x=0\}\), \(\Gamma_{\mathrm{out}}=\{x=L_x\}\).
The inlet flux \(U_{\mathrm{in}}\) implements the chosen perturbation:
$$
U_{\mathrm{in}}(y,t)=
\begin{cases}
1 + \epsilon \cos\!\big(2\pi(y-L_y/2)/L_y\big) & \text{for } t<t_{\mathrm{pert}} \ \text{(sinusoidal)},\\[2pt]
1 + \epsilon\,\text{rand}(y) & \text{for }t<t_{\mathrm{pert}} \ \text{(random)},\\[2pt]
1 & \text{for }t\ge t_{\mathrm{pert}},
\end{cases}
$$
where the random profile \(\text{rand}(y)\) is a smooth, zero–mean function on \([0,L_y]\) (e.g. a truncated Fourier series with random coefficients). The velocity is then
\(\boldsymbol{u}^n = -\,m(T^{n-1})\,\nabla p^n\).

\item \textbf{Temperature solve with stabilizations.} Find \(T^n\in S\) such that for all \(b\in S\),
\begin{align}
&\frac{1}{\Delta t_{\!*}}\int_\Omega T^n b\,\mathrm{d}x
\;+\;\underbrace{\frac{1}{2}\int_\Omega\!\Big[(\boldsymbol{u}^n\!\cdot\nabla T^n)b
-(\boldsymbol{u}^n\!\cdot\nabla b)\,T^n\Big]\mathrm{d}x}_{\text{skew-symmetric advection}}
\;+\;\frac{1}{\Peclet}\int_\Omega \nabla T^n\cdot\nabla b\,\mathrm{d}x\notag\\
&\quad+\;\underbrace{\int_\Omega \big(\mathbf{K}_\alpha^n\nabla T^n\big)\cdot\nabla b\,\mathrm{d}x}_{\text{ramped cross-diffusion}}
\;+\;\Gamma\int_\Omega T^n b\,\mathrm{d}x\notag\\
&\quad+\;\underbrace{\int_\Omega \tau\,(\boldsymbol{u}^n\!\cdot\nabla b)\,R(T^n)\,\mathrm{d}x}_{\text{SUPG}}
\;
=\;\frac{1}{\Delta t_{\!*}}\int_\Omega T^{n-1} b\,\mathrm{d}x,
\label{eq:fem_T_smallGamma}
\end{align}

with Dirichlet condition at the inlet $T^n(0,y) = 1$ and periodic MPC on \(y=0,L_y\). Here:
\begin{itemize}
\item \(\Delta t_{\!*}\) is a ramped time step:
$\Delta t_{\!*}=\Delta t_0<\Delta t$ for the first few steps (startup), then $\Delta t_{\!*}=\Delta t$.
\item The cross–diffusion tensor is
$$
\mathbf{K}_\alpha^n
=\alpha_n\,\frac{2\Peclet}{105}\,\boldsymbol{u}^n\otimes\boldsymbol{u}^n,
$$
where the coefficient $\alpha_n$ increases monotonically from 0 to 1 over the first few steps.
\item The SUPG (Streamline Upwind Petrov–Galerkin) residual and intrinsic time scale are
$$
R(T^n)=\frac{T^n-T^{n-1}}{\Delta t_{\!*}}
+\boldsymbol{u}^n\!\cdot\nabla T^n
-\frac{1}{\Peclet}\,\nabla\!\cdot(\nabla T^n)
-\nabla\!\cdot(\mathbf{K}_\alpha^n\nabla T^n)
+\Gamma\,T^n,
$$
$$
\tau=\left[\left(\frac{2\|\boldsymbol{u}^n\|}{h}\right)^2
+\left(C_k\,\frac{1/\Peclet}{h^2}\right)^2
+\left(C_g\,\Gamma\right)^2\right]^{-\tfrac12},
$$
with cell size \(h\), and \(C_k,C_g=O(1)\).
\end{itemize}
After each solve we clip \(T^n\) to \([0,1]\) to suppress tiny overshoots caused by advection–reaction balancing.
\item \textbf{Advance.} If \(t^n\ge t_{\mathrm{tot}}\) stop, else increase \(n\rightarrow n+1\) and repeat from step (ii).
\end{enumerate}

Some remarks on the algorithm stability and implementation:
\begin{enumerate}
    \item The skew–symmetric advection in \eqref{eq:fem_T_smallGamma} yields a non-dissipative centered discretization that does not amplify in the \(L^2\) norm.
    \item The SUPG (Streamline Upwind Petrov–Galerkin) term stabilizes the advection–reaction operator uniformly in \(\|\boldsymbol{u}\|\) and \(\Gamma\), with \(\tau\) blending advective, diffusive and reactive scales.
    \item Ramping \(\alpha_n\) and \(\Delta t_{\!*}\) over the first \(\mathcal{O}(10)\) steps damps start-up transients, which are more pronounced when the base state has a long decay tail (small \(\Gamma\)).
    \item For the pressure solve \eqref{eq:fem_p_smallGamma} we use PETSc CG with Hypre BoomerAMG \citep{yang2002boomeramg}; for \eqref{eq:fem_T_smallGamma} we use GMRES with a parallel preconditioner (additive Schwarz with ILU(1) on subdomains). Typical PETSc options are:
\begin{align*}
&\texttt{ksp\_type=gmres},\quad \texttt{ksp\_rtol}=10^{-10},\quad \texttt{ksp\_gmres\_restart}=50,\\
&\texttt{pc\_type=asm},\quad \texttt{sub\_pc\_type=ilu}.
\end{align*}
    \item Inlet perturbations of velocity are switched off at \(t_{\mathrm{pert}}\) by setting the corresponding amplitudes to zero; the MPC periodicity ensures the transverse mean is preserved.
    \item We monitor global flux balance \(\sum_{\partial\Omega}\int \boldsymbol{u}\!\cdot\!\boldsymbol{n}\,ds\approx0\) to verify the mixed formulation and boundary conditions.
\end{enumerate}
All simulations are implemented in FEniCSx\,0.9.0 \citep{baratta2023dolfinx} with \texttt{dolfinx\_mpc}. The implementation can be found in the Git repository \citep{CoolingGit2025} in the file \texttt{cooling.py}.

\subsection{Finite element scheme for the linearized model}
\label{sec:num_meth_lin}
To solve the linearized model \eqref{eq:lsa_k} we use a first-order implicit time-stepping scheme with mixed finite elements.
The scheme can be written as the following:
\begin{enumerate}[label=(\roman*)]
    \item Initialize $T^0(x)$ according to \eqref{eq:base_state}. Set $n=1$.
    \item Find $(T^n, u^n) \in W$ such that for all $(Q, v) \in W$:
\begin{subequations}
    \begin{multline}
    \frac{1}{\Delta t} \int_0^{L_x} (T^n - T^{n-1}) Q \, \diff x 
        + \int_0^{L_x} \d {T^n} x Q \, \diff x 
        + \kappa_{\rm eff} \int_0^{L_x} \d {T^n} x \d Q x \, \diff x \\
        + (\kappa k^2 + \Gamma) \int_0^{L_x} {T^n} Q \, \diff x
        + \xi \int_0^{L_x} T_0(x) \left[ -(1 + 2 \kappa_{\parallel} \xi) u^n + \kappa_\parallel \d {u^n} x \right] Q \, \diff x = 0
\end{multline}
and
\begin{align}
    - \int_0^{L_x} \d {u^n} x \d v x \, \diff x 
        + \psi \xi \int_0^{L_x} T_0(x) \d {u^n} x v \, \diff x
        - k^2 \int_0^{L_x} u^n v \, \diff x 
        + k^2 \psi \int_0^{L_x} T^n v \, \diff x = 0
\end{align}\label{eq:fem_lin}\end{subequations}
under the Dirichlet boundary conditions $T^n (0) = 0$ and $u^n(0) = \epsilon \Theta(t_{\rm pert} - t)$ where $\Theta$ is the Heaviside step function.
\item If $n \Delta t \geq t_{\rm tot}$, where $t_{\rm tot}$ is the total simulation time, stop the simulation. Otherwise, increase $n \rightarrow n + 1$ and go to step (ii).
\end{enumerate}

The equation system \eqref{eq:fem_lin} is solved using FEniCS 2019.2.0 \citep{logg2012automated} by means of a direct solver (LU factorization).
The implementation can be found in the Git repository \citep{CoolingGit2025} in the file \texttt{linear\char`_model.py}.

\subsection{Mesh length and resolution}

To run simulations of the linearized model, we must specify the length of the one-dimensional domain $L_x$, the number of spatial tiles $N_x$, the time step $\delta t$, and the total simulation time $t_{\rm tot}$.

A rough estimate of the spatial extent of the solution can be obtained from the inverse of the characteristic length of the base state, $\xi$, defined in Eq. \eqref{eq:def_xi}. For any given cooling rate $\Gamma$ and Péclet number $\Peclet$, the value of $\xi$ follows directly from Eq. \eqref{eq:def_xi}. We therefore set the domain size to $L_x = 10/\xi$. Since the maximum of $T_x$ lies near the inlet of the domain (see Fig. \ref{fig:T_and_u_vs_x}), the ratio between its peak value and its value at the far boundary $x = L_x$ can be estimated as $\max_x \hat T_k(x) / T_k(L_x) \simeq e^{10}$. This indicates that the tail of the temperature perturbation decays by roughly four orders of magnitude over the chosen domain length, ensuring that boundary effects at $x = L_x$ remain negligible.

Similarly, because the characteristic timescale for the instability to grow is of order $\Gamma^{-1}$, we set the total simulation time to $t_{\rm tot} = 10/\Gamma$, which guarantees that the evolution spans at least four order of magnitudes, sufficient for measuring $\gamma$ via a best fit procedure (see Fig. \ref{fig:T_and_u_vs_t}).

We use $N_x = 2000$ spatial tiles and choose the time step as $\delta t = t_{\rm tot}/5000$. Tests with finer spatial meshes and smaller time steps showed no significant change in the computed growth rate. The perturbation was applied over a short interval of a few time steps, $t_{\rm pert} = 5\,\delta t$. Varying this prefactor did not affect the results.

For the full non-linear simulations, we also need to specify the length of the domain along the $y$-axis, $L_y$, and the number of rectangular tiles, $N_y$, along it. For the full simulations shown in Section \ref{sec:compare}, knowing from LSA that $k_{\max}\propto\Gamma$, we set $L_y = 20/\Gamma$ such that the error $\Delta k = 2\pi/L_y$ scaled with $\Gamma$, as visible in Fig. \ref{fig:max_values_vs_Pe_compare}. We also set $L_x = 10/\Gamma$, $N_x = 200$, $N_y = 1000$ and same time length and resolution of the simulations from the linearized model. This choice allowed to obtain a good agreement between the measured $\gamma_*$ ($k_*$) and the corresponding $\gamma_{\max}$ ($k_{\max}$) from LSA. For a proper study of the non-linear terms of the solution, a more refined choice is possibly needed. However, this goes beyond the scope of the our present work.

The linearized model was run on a single node with 4× Intel® Xeon® Gold 6254 CPUs (3.10 GHz, 72 physical cores / 144 threads total) and typically completed in 2–3 minutes. To reconstruct the dispersion relationship and determine ($k_{\max}$, $\gamma_{\max}$), we perform multiple simulations at different $k$-values, thus requiring a few hours of computation time for each combination of global parameters explored. In contrast, full non-linear simulations were executed using 64 MPI processes and took several hours each on average. Depending on the scheduler configuration, these parallel simulations may have utilized one or more nodes.

\section{Large Péclet limit of the linearized model}
\label{sec:app:largePe}
We consider the linearized equations \eqref{eq:lsa_hatk}, inserting $T_0 = e^{-\xi x}$:
\begin{subequations}
\begin{align}
    \left( \dd{}{x} + \psi \xi e^{-\xi x} \d{}{x} - k^2\right) \hat u_k  &=  - k^2 \psi \hat T_k, \label{eq:darcy_laplace_simp} \\
    \left(\gamma + \d{}{x} - \kappa_{\rm eff} \dd{}{x} + \kappa k^2 + \Gamma \right) \hat T_k &= \xi e^{-\xi x} \left( 1 + \kappa_{\parallel}  \left(2 \xi - \d{}{x} \right) \right) \hat u_k \label{eq:advdiff_laplace_simp}.
\end{align}
\label{eq:linmod_simp}
\end{subequations}
We use the length scale $\xi^{-1}$ to rescale $x$ and $k$:
\begin{align}
    x = \xi^{-1} \tilde x, \quad k = \xi \tilde k.
    \label{eq:xk_scaling}
\end{align}
Equation \eqref{eq:linmod_simp} then becomes:
\begin{subequations}
\begin{align}
    \left( \dd{}{\tilde x} + \psi e^{-\tilde x} \d{}{\tilde x} - \tilde k^2\right) \hat u_k  &=  -\tilde k^2 \hat \tau_k, \label{eq:darcy_laplace_simp2} \\
    \left(\gamma +\xi \d{}{\tilde x} - \kappa_{\rm eff} \xi^2 \dd{}{\tilde x} + \kappa \xi^2 \tilde k^2 + \Gamma \right) \hat \tau_k &= \xi \psi e^{-\tilde x} \left( 1 + \kappa_{\parallel} \xi \left(2 - \d{}{\tilde x} \right) \right) \hat u_k \label{eq:advdiff_laplace_simp2},
\end{align}\label{eq:linmod_simp2}\end{subequations}
where we defined $\hat\tau_k = \psi \hat T_k$.

We now assume the high $\Peclet$ limit, which means $\kappa_{\rm eff}\simeq \kappa_\parallel$. Recalling from \eqref{eq:def_xi}, remembering that $\Peclet\, \Gamma \ll 1$, this implies
\begin{align}
    \xi \simeq \frac{- 1 + \sqrt{1 + 4\Gamma \kappa_{\rm \parallel}}}{2 \kappa_{\rm \parallel}}
    \simeq \Gamma.
    \label{eq:xi_simp}
\end{align}
Combining these assumptions with \eqref{eq:linmod_simp2} and \eqref{eq:xi_simp}, Eq. \eqref{eq:darcy_laplace_simp2} remains unchanged while \eqref{eq:advdiff_laplace_simp2} simplifies to
\begin{align}
    \left(\frac{\gamma}{\Gamma} + 1 + \d{}{\tilde x} \right) \hat \tau_k &= \psi e^{-\tilde x}  \hat u_k
    \label{eq:advdiff_laplace_highPe}.
\end{align}
Since $\kappa \Gamma = \Gamma/\Peclet\ll 1/\Peclet^2 \ll 1$. We see from \eqref{eq:darcy_laplace_simp2} and \eqref{eq:advdiff_laplace_highPe} that the only dimensionless parameters are $\tilde\gamma \equiv \gamma/\Gamma$, $\tilde k$ and $\psi$.
Hence, the growth rate $\gamma$ must in the high $\Peclet$ limit be given by
\begin{align}
    \gamma ( k; \Gamma, \kappa_{\parallel}, \psi) = \Gamma \tilde \gamma (\tilde k; \psi).
\end{align}
Since the rescaled dispersion relationship $\tilde\gamma(\tilde k)$ depends only on the parameter $\psi$, its global maximum and the corresponding wavelength, which we call $\tilde \gamma_\textrm{max}$ and $\tilde k_\textrm{max}$ respectively, are solely functions of $\psi$ as well. Hence,
\begin{subequations}
\begin{align}
    k_{\rm max} ( \kappa_\parallel,\Gamma, \psi ) &= \Gamma \tilde k_{\rm max} (\psi), \\
    \gamma_\textrm{max} ( \kappa_\parallel, \Gamma, \psi) &= \gamma(k_\textrm{max}; \kappa_\parallel, \Gamma, \psi) = \Gamma \tilde \gamma (\tilde k_\textrm{max} (\psi); \psi) = \Gamma \tilde \gamma_{\rm max} (\psi),
\end{align}\label{eq:supp_kmax_and_gammamax_largePe}\end{subequations}
In summary, as $\Peclet \gg 1$, the asymptotic scalings for the fastest mode and its growth rate are $k_{\rm max} \sim \Gamma$ and $\gamma_{\rm max} \sim \Gamma$, respectively.
We further note that as $\psi_c$ is the value of $\psi$ where $\gamma_\textrm{max} (\psi_c) = 0$, it follows that for high Péclet numbers $\beta_c=e^{-\psi_c}$ is a universal numerical value independent of any system parameters.

\subsection{Large viscosity ratio}
\label{sec:app:largePe_smallbeta}
Taking the analysis a step further, we may assume that $\psi$ is \emph{large}.
We introduce the scaled quantities $\tilde{\tilde k} = \psi^{-1} \tilde k$ and $\tilde{\tilde x} = \psi \tilde x$.
Equations \eqref{eq:darcy_laplace_simp2} and \eqref{eq:advdiff_laplace_highPe} become
\begin{subequations}
\begin{align}
    \left( \dd{}{\tilde{\tilde x}} + e^{-\tilde{\tilde x}/\psi} \d{}{\tilde{\tilde x}} - \tilde{\tilde k}^2\right) \hat u_k  &=  -\tilde{\tilde k}^2 \hat \tau_k, \label{eq:darcy_laplace_psilarge} \\
    \left( \frac{\frac{\gamma}{\Gamma} + 1}{\psi} + \d{}{\tilde{\tilde x}} \right) \hat \tau_k &= e^{-\tilde{\tilde x}/\psi} \hat u_k \label{eq:advdiff_laplace_highPe_psilarge}.
\end{align}\label{eq:model_highPe_psilarge}\end{subequations}
$\hat u_k$ will typically decay with $x$ as $\hat u_k \sim e^{-\tilde \eta \tilde x}$ where $\tilde \eta = \min \{ \tilde k, \tilde \gamma \}$.
Hence, for large $\psi$, $e^{-\tilde{\tilde x}/\psi} \simeq 1$ wherever $\hat u_k$ is significant.
To the lowest order in the small parameter $1/\psi$, equations \eqref{eq:model_highPe_psilarge} become
\begin{subequations}
\begin{align}
    \left( \dd{}{\tilde{\tilde x}} + \d{}{\tilde{\tilde x}} - \tilde{\tilde k}^2\right) \hat u_k  &=  -\tilde{\tilde k}^2 \hat \tau_k, \label{eq:darcy_laplace_psilarge2} \\
    \left( \frac{\frac{\gamma}{\Gamma} + 1}{\psi} + \d{}{\tilde{\tilde x}} \right) \hat \tau_k &= \hat u_k \label{eq:advdiff_laplace_highPe_psilarge2}.
\end{align}\label{eq:model_highPe_psilarge2}\end{subequations}
Here, all the system parameters are grouped together in the first term of \eqref{eq:advdiff_laplace_highPe_psilarge2}, which should be solely given by $\tilde{\tilde k}$.
This scaling thus explains why
$$k_\textrm{max} \sim \psi, \quad \gamma_{\max}/\Gamma + 1 \sim \psi$$ 
for large $\psi$, up to a numerical prefactor.

It should be noted that expanding \eqref{eq:model_highPe_psilarge} to the first order in $1/\psi$ would give a more accurate estimate of $k_\textrm{max} \sim \psi ( 1 + \textrm{const.}/\psi ) \sim \textrm{const.} + \psi$ (and similar for $\gamma_\textrm{max}$). Nevertheless, this linear functional form is in good agreement with our data for $\psi > \psi_c$.

\bibliographystyle{apsrev4-2}
\bibliography{references}

\end{document}